\documentclass[a4paper,fleqn,usenatbib]{mnras}

\usepackage[T1]{fontenc}
\usepackage{ae,aecompl}

\usepackage{graphicx}	% Including figure files
\usepackage{amsmath}	% Advanced maths commands
\usepackage{amssymb}	% Extra maths symbols
 
\title[Sizes and Angular Momenta Evolution]{The Evolution of Sizes and Specific Angular Momenta in Hierarchical Models of Galaxy Formation and Evolution}

\author[Zoldan et al.]{
Anna Zoldan,$^{1}$\thanks{E-mail: anna.zoldan@inaf.it}
Gabriella De Lucia,$^{1}$
Lizhi Xie,$^{2}$
Fabio Fontanot$^{1}$ and 
\newauthor
Michaela Hirschmann$^{3}$
\\
$^{1}$OATS, INAF, Via Bazzoni 2, 34124-Trieste, TS, Italy\\
$^{2}$Tianjin Astrophysics Center, Tianjin Normal University, Binshuixidao 393, 300384, Tianjin, China\\
$^{3}$Institut d'Astrophysique de Paris, Sorbonne Universit\'{e}s, UPM-CNRS, UMR7095, F-75014, Paris, France
}

\date{Accepted XXX. Received YYY; in original form ZZZ}

\pubyear{2017}

\begin{document}
\label{firstpage}
\pagerange{\pageref{firstpage}--\pageref{lastpage}}
\maketitle

% Abstract of the paper
\begin{abstract}
We extend our previous work focused at $z\sim0$, studying the redshift evolution of galaxy dynamical properties using the state-of-the-art semi-analytic model GAEA:
we show that the predicted size-mass relation for disky/star forming and quiescent galaxies is in good agreement with observational estimates, up to $z\sim2$. 
Bulge dominated galaxies have sizes that are offset low with respect to observational estimates, mainly due to our implementation of disk instability at high redshift. 
At large masses, both quiescent and bulge dominated galaxies have sizes smaller than observed. 
We interpret this as a consequence of our most massive galaxies having larger gas masses than observed, and therefore being more affected by
dissipation. 
We argue that a proper treatment of quasar driven winds is needed to alleviate this problem. 
Our model compact galaxies have number densities in agreement with observational estimates and they form most of their stars in small and low
angular momentum high-$z$ halos. 
GAEA predicts that a significant fraction of compact galaxies forming at high-$z$ is bound to merge with larger structures at lower redshifts: therefore they are not the progenitors of normal-size passive galaxies at $z=0$. 
Our model also predicts a stellar-halo size relation that is in good agreement with observational estimates. 
The ratio between stellar size and halo size is proportional to the halo spin and does not depend on stellar mass but for the most massive galaxies, where AGN feedback leads to a significant decrease of the retention factor (from about 80 per cent to 20 per cent).

\end{abstract}

% Select between one and six entries from the list of approved keywords.
% Don't make up new ones.

%%%%%%%%%%%%%%%%%%%%%%%%%%%%%%%%%%%%%%%%%%%%%%%%%%

%%%%%%%%%%%%%%%%% BODY OF PAPER %%%%%%%%%%%%%%%%%%

%------------------------------------------------
%------------------INTRODUCTION-------------------
%------------------------------------------------
\section{Introduction}
The energy of a galaxy can be evaluated through the simultaneous measurement of its mass, radius and angular momentum. 
These information can be used to quantify the energy radiated during galaxy formation, and to estimate the dissipative processes at play.
The link between galaxy and halo properties can be explained through the simple model described in \citet{fall1980intro} and \citet{mo1998DI}. 
In the current standard picture of structure formation, gas is believed to condense  at the center of dark matter (DM) halos, forming galaxies. 
The tidal torques that arise during the growth of perturbations give origin to most of the angular momentum of these galaxies \citep{peebles1969,white1984,barnes1987}.
Angular momentum is usually measured through the spin parameter \citep{peebles1969}:
\begin{equation}
 \lambda = \frac{J |E|^{1/2}}{G M^{5/2}}
\end{equation}
where $J$ is the angular momentum, $E$ the internal energy, $M$ the mass of the halo, and $G$ the gravitational constant. 
% The angular momentum of the DM halo has a strong dependence on its mass: $j_h\propto M_{h}^{5/3}$  \citep{catelan1996}. 
The gas is assumed to share the dynamical properties of the DM halo (at first order, we ignore self gravity), and to preserve its specific angular momentum ($j_h$) during gas cooling.
For a halo with a spherical isothermal density profile ($\rho \propto r^{-2}$), and  assuming the gas collapses in a disk with exponential surface density profile, its scale radius can be expressed as $R_d = \frac{1}{\sqrt{2}} \lambda R_h$, where $R_h$ is the radius of the DM halo. 
The model can be refined including e.g. deviations from the isothermal profile, modifications of the inner halo profile due to self-gravity or energy input by stellar feedback or active galactic nuclei, angular momentum transfer during disk formation or mergers \citep{blumenthal1986rrh, mo1998DI, dutton2007rrh, somerville2008rrh,shankar2013r_z, porter2014dissipation}.
These improvements still lead to a disk scale radius that is proportional to $\propto \lambda R_h$.

Late-type galaxies (star forming, disky and dynamically young) are those that better preserve the halo dynamical properties and are better described by the theoretical model illustrated above. 
Early-type galaxies, in contrast, have usually a very small retention factor, i.e. they retain a small fraction of the angular momentum of the hosting halo, likely due to events like galaxy mergers and disk instabilities. 

Recent high resolution imaging in multiple photometric bands and Integral Field Spectroscopy allowed measurements of spatially resolved properties for thousands of galaxies (e.g. the Sloan Digital Sky Survey, SDSS- \citealt{york2000sdss}; the Galaxy And Mass Assembly survey, GAMA - \citealt{driver2011gama}; Calar Alto Legacy Integral Field Area survey, CALIFA - \citealt{sanchez2012califa}; the Sydney-Australian-Astronomical-Observatory Multi-object Integral-Field Spectrograph survey, SAMI - \citealt{bryant2015sami}; and the Mapping Nearby Galaxies at APO survey, MaNGA- \citealt{bundy2015manga}).
The first large statistical analysis of the galaxy size--mass relation in the local Universe has been performed by \citet{shen2003_sdss}, using SDSS data.  
The correlation shows a large scatter that is due in part to galaxy morphology, with late-type galaxies having on average larger characteristic radii than early-type galaxies.  Different late/early-type selections give similar size-mass median relations \citep[e.g.][based on GAMA]{lange2015}. 
A similar dependence on galaxy type has been found for the specific angular momentum ($j_*$) versus galaxy stellar mass ($M_*$) relation \citep{fall1983j,romanowsky2012js,obreschkow2014jmbt,cortese2016sami}. 
% ET galaxies exhibit a complex internal dynamical structure, are classified as `fast' or `slow' rotators according to the predominance of the rotational velocity compared to the velocity dispersion, respectively \citep[e.g. ATLAS$^{3D}$][]{cappellari2011atlas3d}.

These studies have also been extended to higher redshift.
\citet{van_der_wel2014candles} have analyzed 30958 galaxies from CANDELS \citep{grogin2011CANDELS} with $M_*>10^{9}\,{\rm M_{\sun}}$ and $0<z<3$, and classified them according to their color. 
The dependence of the half mass radius $R_{1/2}$--mass relation on galaxy type is preserved at high-$z$, with late-type galaxies having a slope shallower than early-type galaxies ($R_{1/2}^{LT}\propto M_*^{0.22}$ and $R_{1/2}^{ET}\propto M_*^{0.75}$, respectively). 
At fixed stellar mass, high-redshift galaxies are smaller than present day ones: at $z\sim2$, late and early-type galaxies are $\sim 2$ and $\sim4$ times smaller than their present-day counterparts, respectively. 
A number of studies have confirmed these trends, and analyzed the origin of the observed size--mass relation \citep{ichikawa2012,cassata2013goods_candles_r_z}.

High redshift and small size early-type galaxies are typically referred to as ``compact'' or even ``ultra-compact'' galaxies, depending on their size. 
The fate of these compact galaxies is still matter of debate, in particular how they contribute to the formation of the present day passive population.
A possible interpretation of their number density evolution is that the process creating and refueling the compact population has stopped to be predominant around $z\sim1$, allowing ``normal'' size galaxies to dominate today \citep{cassata2013goods_candles_r_z,gargiulo2017r_z}.
Theoretically, it has been argued that compact galaxies can evolve into normal present-day galaxies thanks to subsequent dry mergers \citep{naab2009r_z},  and/or to AGN feedback \citep[][]{dubois2013BH_r, choi2018BH_r}. 
An unanimous interpretation of the evolution of compact galaxies is, however, still lacking.

While galaxy sizes have been widely studied up to high redshift for a large variety of galaxy types, stellar specific angular momenta are difficult to measure outside the local Universe. 
At low redshift, the $j_*$--$M_*$ relation resembles the size--mass relation, in particular for its dependence on the galaxy morphology and star formation activity. 
\citet{swinbank2017js_z} and \citet{alcorn2018j_z} measured $j_*$ of star forming galaxies at $z\sim 0.28-1.65$ and $z\sim 1.7-2.5$, respectively, finding little evolution. 
With the available instruments, measurements of $j_*$ for early-type galaxies at high redshift are still unfeasible.

A number of recent studies have focused on the relation between the sizes of galaxies and those of their hosting halos.
\citet{kravtsov2013r50_r200} analyzed the $R_{1/2}$--$R_{200}$ relation at $z=0$, where $R_{200}$ is the radius that encloses an halo over-density larger 200 times the critical density of the Universe. 
He evaluated halo masses and sizes applying an abundance matching technique to data from several publicly available data samples. 
The obtained relation shows a direct proportionality: $R_{1/2}=0.015R_{200}$, in agreement with expectations from models that assume galaxy sizes are controlled by halo angular momentum.
\citet{huang2017r50_r200} have extended these studies up to $z\sim3$ using CANDELS data, and have shown that $R_{1/2}\sim 0.03 R_{200}$ for $0<z<3$, with a negative curvature at the high $R_{200}$ end, due to the predominance of small early-type galaxies in these large halos. 
Late and early-type galaxies stay on distinct, almost parallel, relations, with late-type galaxies showing the strongest correlations with halo sizes (as expected).
% \citet{somerville2018r50_r200} adopted a more sophisticated abundance matching approach on GAMA and CANDELS galaxies to study the dependence of $R_{1/2}/R_{\rm vir}$ and $R_{1/2}/(\lambda R_{\rm vir})$ on stellar mass at various redshifts. 
% They found these quantities are almost independent on stellar mass at low redshift, while at high redshift they are higher at low masses. 

Early theoretical work based on semi-analytic techniques used the model by \citet{mo1998DI} \citep[see][]{kauffmann1996sam,somerville1999sam,cole2000,croton2006}.
Recent work has introduced a more accurate tracing of the angular momentum of galactic components 
\citep[e.g.][]{guo10,xie2017sam,stevens2016darksage}.
Several studies  analyzed the size--mass relation and its evolution  in semi-analytic models \citep{somerville2008sam,shankar2013r_z,tonini2016}  and in hydrodynamical simulations \citep{furlong2015eagle,remus2017magneticum}. 
An important result is that gas dissipation during mergers is a fundamental ingredient for a realistic treatment of early-type galaxies \citep{shankar2013r_z,porter2014dissipation,tonini2016}.
Similar analyses were performed for the specific angular momentum versus mass relation \citep{stevens2016darksage, lagos2017eagle_j, zoldan2018size_j}.
Numerical simulations have shown that sizes and specific angular momenta of galaxies are sensitive to the prescriptions adopted for stellar feedback. 
In particular, strong feedback at high redshift removes low angular momentum gas, producing galaxies with sizes and angular momenta similar to those observed in the Local Universe \citep{ubler2014j_fs,marinacci2014,teklu2015magneticum_j,pedrosa2015sim,genel2015j_illustris}.

In a previous work, we have used a state-of-the-art semi-analytic model to study galaxy dynamical properties such as sizes and angular momenta at redshift $z=0$ \citep{zoldan2018size_j}, finding a good agreement with observational data. 
In this work, we use a slightly updated version of the same model to explore the evolution of galaxy stellar sizes, and study the dependence of sizes and angular momenta on stellar mass, galaxy type and DM halo properties.

The paper is organized as follows.
In section~\ref{sec:model_description}, we review the model and simulations used in this study.
In section~\ref{sec:size_mass_z}, we show the evolution of the size--mass relation obtained using our model, we compare it to observational results, and we study its dependence on other galactic properties.
In section \ref{sec:j_z}, we perform a similar analysis on the $j_*$--$M_*$ relation. 
In section~\ref{sec:r_n_z}, we study the evolution of average size and number density of early and late-type galaxies, focusing on compact galaxies. 
In section~\ref{sec:r_halo_z}, we analyze the dependence of the half-mass radius of model galaxies on halo properties at different cosmic epochs. 
In section~\ref{sec:discussion}, we discuss our results, and we summarize our conclusions in section~\ref{sec:conclusions}.

%------------------------------------------------
%------------------THE MODEL-------------------
%------------------------------------------------
\section{The model}
\label{sec:model_description}
In this work, we use an updated version of the GAlaxy Evolution and Assembly (GAEA) semi-analytic model, described in \citet{hirschmann2015}. 
This model descends from that originally published in \citet{delucia07}, but many prescriptions have been updated significantly since then.  
In particular, GAEA includes a sophisticated treatment for the non instantaneous recycling of gas, metals, and energy \citep{delucia2014}, and a stellar feedback
scheme partly based on results of hydrodynamical simulations \citep{hirschmann2015}.  
In this work, we use the \citet{xie2017sam} GAEA model version, including a specific treatment
for angular momentum exchanges between galactic components, and
prescriptions to partition the cold gas into its molecular (star
forming) and atomic components. 
Specifically, we use the prescription to estimate the molecular gas fraction by \citet{blitz2006}. 
We also include the prescriptions for gas dissipation during major mergers detailed in \citet{zoldan2018size_j}; this treatment is necessary to obtain realistic bulge sizes from gas rich mergers at redshift 0. 

Our fiducial model is able to reproduce a number of important observational measurements: 
the evolution of the galaxy stellar mass function up to $z\sim7$ and of the cosmic star formation rate density up to $z\sim10$ \citep{fontanot2017gaea}; 
the measured correlation between stellar mass/luminosity and metal content of galaxies in the local Universe, down to the scale of Milky Way satellites
\citep{delucia2014,hirschmann2015}, and the evolution of the galaxy
mass--gas metallicity relation up to redshift $z\sim2$
\citep{hirschmann2015,xie2017sam}; 
the size--mass relation and the specific angular momentum--mass relation of late and early-type galaxies in the local Universe \citep{zoldan2018size_j}; 
the quiescent satellite fractions at low redshift \citep{delucia2019satQ}. 

The model is not without problems, however.
Our model massive galaxies tend to form stars at higher rates than observed \citep{hirschmann2015}, and the model tends to under-predict the measured level of star formation activity at high redshift \citep{xie2017sam}.
In addition, the disk instability treatment leads to highly under-sized bulges, in particular for central galaxies with a bulge over total mass ratio $0.5<B/T<0.7$ and mass $10<\log_{10}(M_*\,[{\rm M_{\sun}}])<10.8$ \citep{zoldan2018size_j}. 

In the following, we provide a brief review of those prescriptions in the 
\citet{xie2017sam} model that are relevant for this work.  
For a complete description of our model, we refer to the original papers by \citet{delucia2014}, \citet{hirschmann2015}, and \citet{xie2017sam}.

\subsection{Cosmological simulation and merger trees}
The model outputs used in this work are based on merger trees from the Millennium Simulation \citep[MRI,][]{springel2005}, and from the higher resolution
Millennium II Simulation \citep[MRII,][]{boylan-kolchin2009}.  
The MRI follows the evolution of N=$2160^3$ particles of mass
$8.6\times10^8 h^{-1}{\rm M_{\sun}}$, in a box of $500\;
  h^{-1}$Mpc comoving on a side.  Simulation outputs are stored in 64
snapshots, logarithmically spaced in redshift.  
The MRII corresponds to a simulation box with size of $100\; h^{-1}$Mpc on
a side (corresponding to a volume 125 times smaller than that of the MRI), with a mass resolution that is 125 times better than that used in the MRI.
The cosmological model of the two Millennium simulations is consistent with WMAP1 data \citep{spergel2003WMAP1}, with cosmological parameters $\Omega_{b}=0.045$, $\Omega_{m}=0.25$, $\Omega_{\Lambda}=0.75$, $H_0=100h\;{\rm Mpc^{-1}\;km\;s^{-1}}$, $h=0.73$, $\sigma_{8}=0.9$, and $n=1$.  

Halo merger trees are constructed using a two-step procedure.
First, halos are identified at each snapshot using a classical Friends-of-Friends algorithm, with a linking length equal to 0.2 times the mean inter-particle separation. 
Then, the SUBFIND algorithm \citep{springel2001accr_mode} is used to identify bound
substructures in each FoF halo. 
As in previous work, only substructures with at least 20 bound particles are considered 
genuine: this sets the halo mass resolution to $M_{h}=1.7\times 10^{10}
h^{-1}{\rm M_{\sun}}$ for the MRI and to $M_{\rm h}=1.4\times10^8h^{-1}{\rm M_{\sun}}$ for the MRII.  
A unique descendant is identified for each subhalo in the subsequent snapshot, tracing a subset of the most bound particles \citep{springel2005}. 
In this way, each subhalo is automatically linked to all its progenitors (at all previous snapshots), and its merger history is constructed.  
A main branch for each halo is defined as the one that follows, at each node of the tree, the progenitor with
the largest integrated mass \citep{delucia07}.
For each halo and subhalo, SUBFIND provides estimates of $M_{200}$, $V_{max}$ and $\vec{j_{h}}$, that are the mass corresponding to an over-density of 200 times the critical density of the Universe $\rho_{\rm crit}$, the maximum rotational velocity, and the specific angular momentum, respectively.
A virial radius $R_{200}$ for each halo/subhalo is computed using the formula:
\begin{equation}
\label{eq:r200}
 R_{200} = \left(\frac{M_{200}}{ 200 \cdot 4 \pi /3 \cdot \rho_{\rm crit}} \right)^{1/3}.
\end{equation}

As in previous work, we use the MRII to verify the robustness of our results at masses near the resolution limit of the MRI simulation. 
The resolution limits of the MRI and MRII simulations translate in stellar mass
limits for the X17 model of about $\sim 10^9\;{\rm M_{\sun}}$ and  $\sim 10^8\;{\rm M_{\sun}}$, respectively \citep[see Fig.~6 of ][]{xie2017sam}.
In the following, we will use results from the MRI for stellar masses larger than $10^{10}\,{\rm M_{\sun}}$, and from the MRII for lower masses (down to $10^9\;{\rm M_{\sun}}$).

\subsection{Size and specific angular momentum}
\label{sec:model_size_vmax}
Our model includes an explicit treatment for angular momentum exchanges between galactic components. 
We assume the hot gas reservoir has the same specific angular momentum of the DM halo. 
The cold gas disk acquires angular momentum from the hot gas during cooling, proportionally to the cooled mass. 
The same cold gas specific angular momentum is acquired by newly formed stars, thus the stellar disk acquires angular momentum from the gaseous disk proportionally to the mass of the stars formed.

The scale radii of the cold gas and stellar disks are estimated from their specific angular momenta and rotational velocities.  
Specifically, the disk scale radius is expressed as:
\begin{equation}
 R_{x} = \frac{j_x}{2V_{max}},
\end{equation}
where $R_x$ and $j_x$ are the radius and the specific angular momentum of the $x$-component (either cold gas or stars).  
$V_{max}$ is the maximum rotational velocity of the parent halo.

In our model, the bulge is assumed to be a dispersion dominated spheroid, and its
size is estimated using energy conservation arguments. 
The energies involved in mergers between spheroids are those due to their
gravitational potential and interaction.  
Following results from hydrodynamical simulations \citep{hopkins2009dissipation,covington2011dissipation,porter2014dissipation}, we introduced a treatment for gas dissipation during major mergers \citep{zoldan2018size_j}, based on the formula suggested by \citet{hopkins2009dissipation}.
When a disk instability occurs, the model evaluates the radius enclosing the stellar mass removed from the central part of the disk (to restore the stability), assuming it distributes in a disk with an exponential surface density profile. 
We then use this radius as the scale-radius of the newly formed spheroidal component.
If a bulge already exists, we merge the newly formed spheroid with the pre-existing bulge, assuming energy conservation.

\subsection{Modifications with respect to our previous modeling of galaxy sizes}
In \citet{zoldan2018size_j}, we have shown that the treatment for disk and bulge sizes  described in section~\ref{sec:model_size_vmax} leads to realistic median size--mass relations for both late (LT) and early-type (ET) galaxies at $z=0$, for a selection based on $B/T$. 
In our previous work, the $V_{max}$ used to infer the disk scale radius was evaluated at each snapshot of the simulation, using the outputs from SUBFIND, also for satellite galaxies.
This tends to overestimate satellite radii for subhalos that are poorly resolved (and have under-estimated rotational velocity). 
In this work, we have revised our treatment and assume that, for satellite galaxies, the value of $V_{max}$ remains constant after infall. 
This assumption is supported by the findings that $V_{max}$ is much less affected than subhalo mass by tidal stripping within the potential of the parent halo \citep[e.g.][]{hayashi2003sat_vel,kravtsov2004subhalos}.
In addition, galaxies residing at the center of dark matter subhalos should be much more resilient to the disruption processes that strongly affect their parent halo, because baryons are much more concentrated and their presence deepens the gravitational potential of the suhalos.

%------- Reff vs M* with Reff from the profiles, for model with fixed and changing Vmax; ang_mom/j_integration/plot_test.py (profiles made with calc_more_jaffe.py)
\begin{figure}
    \includegraphics[trim=0cm 0cm 1cm 1cm, clip, width = 0.99\linewidth]{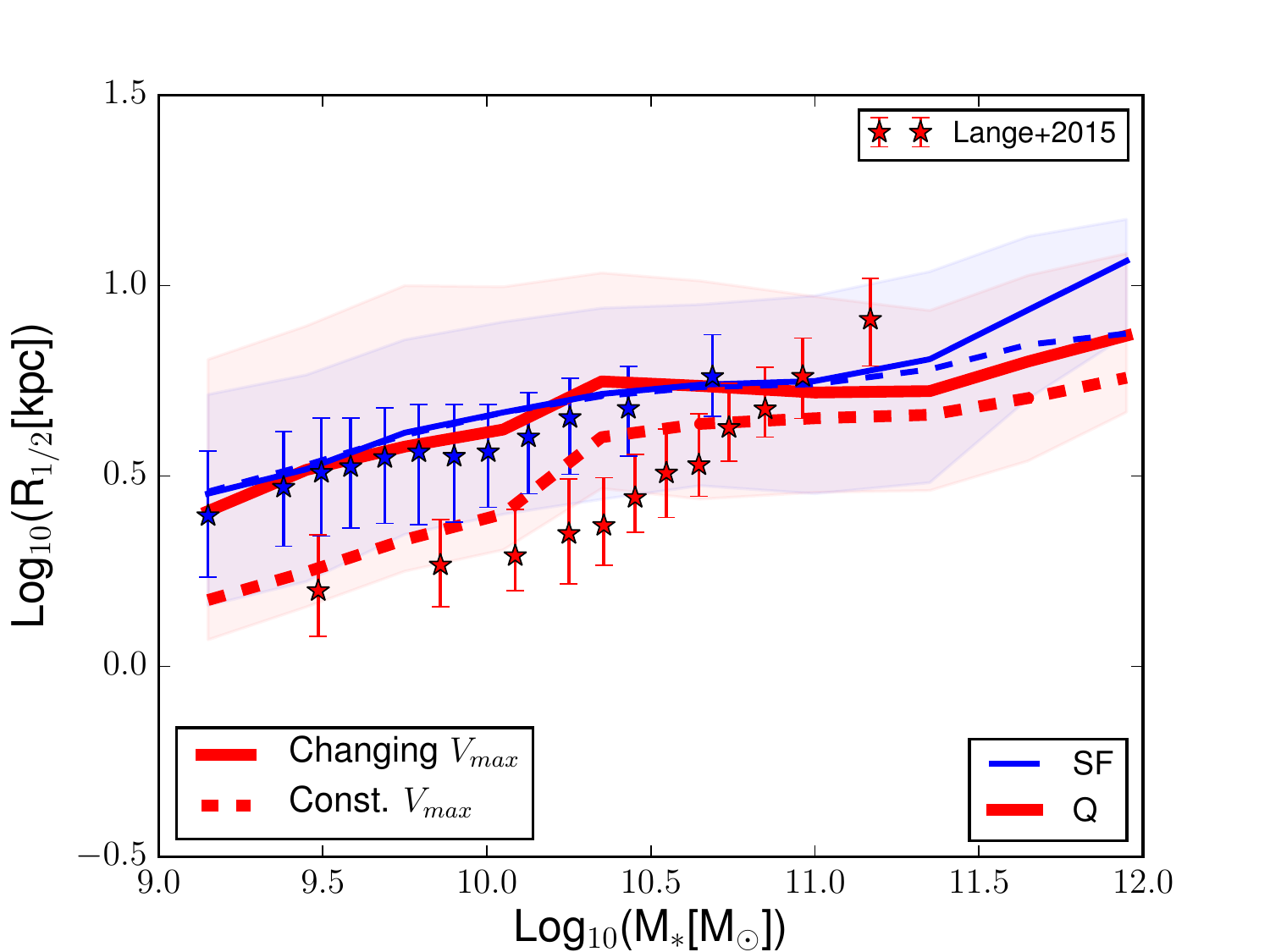}
    \caption{The $R_{1/2}$--$M_*$ relation for SF and Q model galaxies (blue
      and red lines) for the original model (solid lines) and the new version with constant $V_{max}$ after accretion (dashed lines). 
      SF and Q galaxies are selected using a sSFR$=0.3/t_H$ threshold. 
      Shaded areas show the region between the 16th and 84th percentiles of the original model distribution. The scatter of the modified model is not shown for clarity, but its magnitude is similar to that of the older model.
      Observational data by \citet{lange2015} are show, as a reference, as stars with errorbars.}
    \label{fig:reff_ms_noVmax_Vmax}
\end{figure}
This modification does not influence the size-mass relation obtained for LT and ET galaxies selected using a $B/T$ threshold, while in the case of a selection based on a specific Star Formation Rate (sSFR) threshold we find significant differences.
This can be appreciated in figure~\ref{fig:reff_ms_noVmax_Vmax}, where we show the median size--mass relations (lines) for model galaxies classified as star forming (SF, blue) and quiescent (Q, red), using a threshold in sSFR (sSFR$=0.3/t_H$, where $t_H$ is the age of the Universe at each redshift). 
We show the fiducial model as solid lines, and the model with $V_{max}$ fixed after accretion as dashed lines. 
The quantity shown is the half-mass radius of the projected stellar mass, assuming that all model galaxies are projected face-on \citep[as in][]{zoldan2018size_j}. 
We add observational measurements by \citet{lange2015} as stars with errorbars to guide the eye.

In the original model, Q galaxies have sizes similar to those of SF galaxies.
This is due to a large fraction of quiescent, but disky  satellites.
In the modified version of the model, these satellites have smaller sizes, because their $V_{max}$ does not decrease due to the tidal stripping of the halo. 
A constant $V_{max}$ after accretion brings the median size of passive galaxies in better agreement with observations, although model predictions are still somewhat larger than observational data at intermediate stellar masses, and slightly smaller for most massive galaxies. 

Although a constant $V_{max}$ does not affect the size--mass relation for galaxies selected using a $B/T$ threshold, it affects the number of ET satellite galaxies  (there are almost $\sim4$ times more ET satellites than in the original model, for $10<\log_{10}(M_*\,[{\rm M_{\sun}}])<11$).
This occurs because a smaller radius implies a larger probability of having a disk instability, increasing the bulge mass. 

{ \it In the following, we will use the model variant assuming a constant $V_{max}$ after infall to evaluate the size-mass and specific angular momentum-mass evolution. 
This is important, as most of the observations at high redshift are based on a sSFR or color-color classification.}

%------------------------------------------------
%---------------SIZE-MASS RELATION EVOLUTION-----
%------------------------------------------------
\section{The size--mass relation evolution}
\label{sec:size_mass_z}
%------- Reff vs M* with Reff from the profiles, at various redshifts; ang_mom/j_integration/c_func/halo_prop/plot_final_reff_ms.py (profiles made with calc_more_jaffe.py)
\begin{figure*}
    \includegraphics[trim=2cm 0.8cm 2.5cm 2cm, clip, width = 0.99\linewidth]{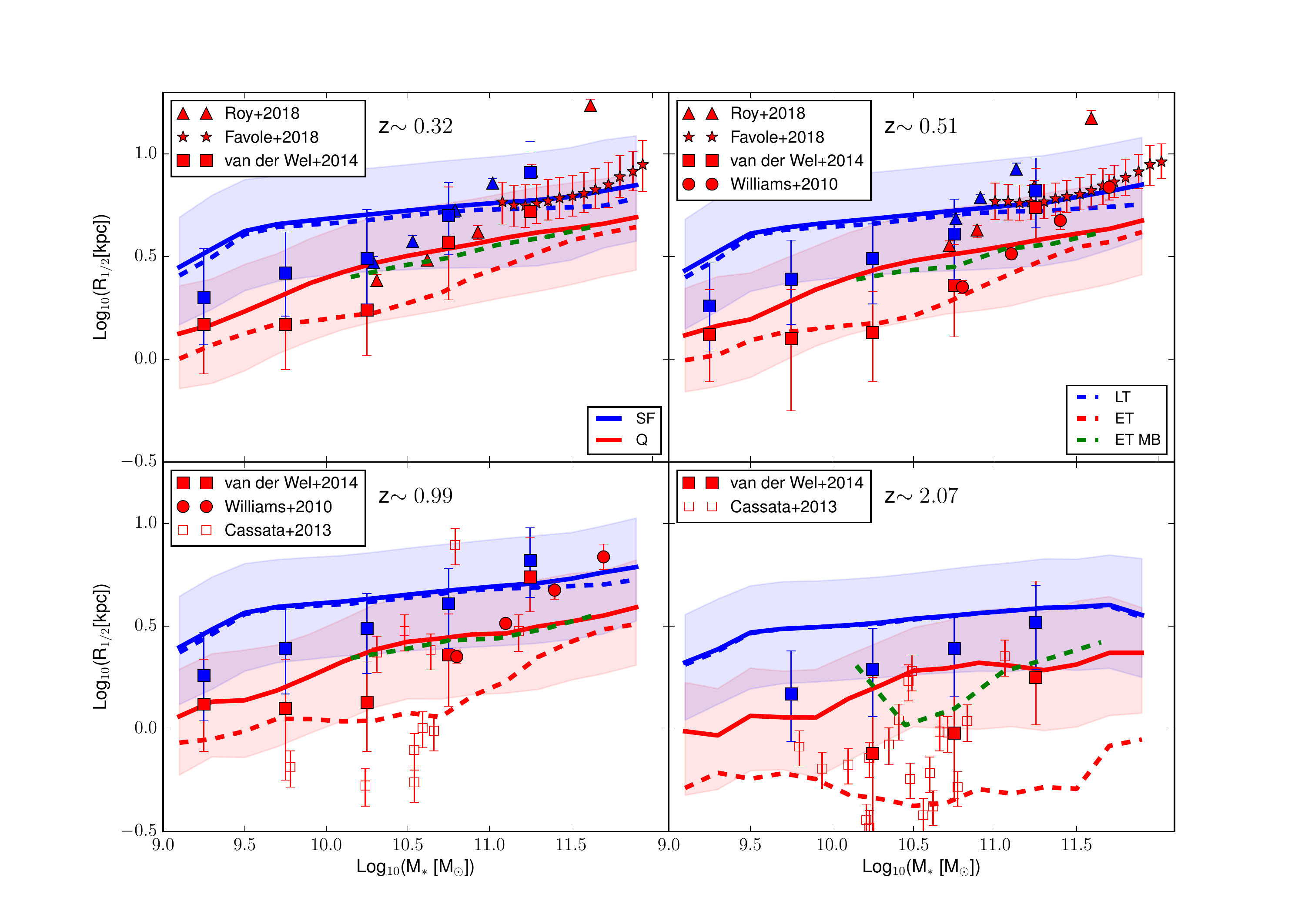}
    \caption{The $R_{1/2}$--$M_*$ relation for selected model galaxies. 
      SF/Q galaxies are selected using a sSFR$=0.3/t_H$  threshold (solid blue/red), while LT and ET galaxies are selected using a $B/T=0.5$ threshold (dashed blue/red lines). 
      Shaded areas show the regions between the 16th and 84th percentiles of the SF/Q distributions, while the scatter in the LT/ET selection is not shown for clarity (the scatter is similar for SF/LT galaxies, while it extends to lower values of $R_{1/2}$ in the case of ET galaxies, in particular in the intermediate mass range).
      Different panels show different redshifts, as indicated in the captions.
      In each panel, we show  the median relation for $B/T>0.5$ galaxies that have a bulge formed mainly through merger (green dashed line, shown only for the MRI). 
      Symbols with error bars correspond to different observational measurements, as indicated in the legend. 
      Observations are selected to be close to the redshift of each panel.}
    \label{fig:reff_ms_redshift}
\end{figure*}
We have analyzed in detail our model predictions for the size--mass relation at redshift $z=0$ in a previous paper \citep{zoldan2018size_j}.
We have found a good agreement with observational data, for both LT and ET galaxies (selected using a $B/T$ threshold). 
We have also shown that bulges formed mainly through disk instabilities have unrealistically small sizes.
This influences, in particular, central galaxies selected using a $B/T=0.5$ threshold, in the mass range $M_*=10^{10.2}-10^{10.8}\,{\rm M_{\sun}}$.

In figure~\ref{fig:reff_ms_redshift}, we show the median size--mass relation at four different redshifts, for selected model galaxies: SF (blue) and Q (red) galaxies (sSFR$>$ and $<0.3/t_H$) are shown as solid lines, while LT (blue) and ET (red) galaxies ($B/T<$ and $>0.5$) are shown as dashed lines.
The 16th-84th percentiles of the distributions of SF/Q galaxies are shown as shaded areas. 
% The scatter of LT galaxies is almost identical to that of SF galaxies, while that of ET galaxies extends to lower sizes in the intermediate stellar mass range ($M_*\in[10^{10};\,10^{11.2}]\,{\rm M_{\sun}}$).
Observational data are described in section~\ref{sec:obs_reff}, and are represented as symbols with errorbars.

We find a reasonable agreement between model predictions and observational estimates, in particular for SF/Q and LT galaxies.
In the next sub-sections, we will discuss the results shown in this figure in more detail.

\subsection{Observational data}
\label{sec:obs_reff}

\citet{williams2010uds_r_z} selected Q galaxies from the near-infrared images of the UKIDSS UDS Data Release 1 and the H-band mosaic images of the UKIDSS Data Release 3 \citep{lawrence2007uds, warren2007uds}. 
They obtained  circularized effective radii by performing S\'ersic fits for all  galaxies with $K<22.4$.
Stellar masses and sSFRs were calculated using SED fitting.
Their galaxies are divided in three bins of photometric redshifts, and quiescent galaxies are selected applying a cut in specific star formation: ${\rm sSFR}<0.3/t_H$, where $t_H$ is the age of the Universe at each redshift. 
They fitted the size--mass relation down to $M_*=10^{10.8}\; {\rm M_{\sun}}$.
% , using the fitting function $\log_{10}(r_{eff}) = A + b\cdot(\log_{10}(M_*/{\rm M_{\sun}}) -11)$.

We show also results by \citet{cassata2013goods_candles_r_z}: they used the GOODS-South field, including measurements from several surveys.
They evaluated stellar masses and sSFRs using SED fitting. 
Q galaxies were selected to have $M_*>10^{10}\;{\rm M_{\sun}}$, $z>1.2$, sSFR$<10^{-2} {\rm Gyr^{-1}}$, a spheroidal shape and no dust emission. 
Only 28 of their 107 galaxies have spectroscopic redhsifts. 
The size corresponds to a circularized half-light radius obtained using S\'ersic fitting. 

\citet{van_der_wel2014candles} analyzed the size-mass relation for SF and Q galaxies up to $z\sim 2.75$, using measurements based on the 3D-HST survey \citep{brammer20123DHST} and CANDELS \citep{grogin2011CANDELS,koekemoer2011candles}. 
SF and Q galaxies were classified according to the rest frame U-V vs V-J color diagram. 
The measured size is the circularized effective radius.

We also consider observational data by \citet{favole2018boss_decals}, who used  DECaLS DR3 survey photometry \citep{comparat2013decals,comparat2016decals} matched to the SDSS-III/BOSS DR12 data \citep{anderson2012boss,bolton2012boss} to infer the size-mass relation  for the most massive  galaxies up to redshift $z\sim0.6$.  
They use circularized effective radii of Luminous Red Galaxies (LRGs). 
On the basis of their S\'ersic index, De Vaucouleurs profiles and colors, these galaxies are classified in large majority as ET/Q galaxies ($\sim 80\%$). 

Finally, we show data by \citet{roy2018kids}, obtained using the Kilo Degree Survey \citep{de_jong2015kids}. 
In this work, the authors provide an estimate of the size--mass relation for disk and spheroid-dominated galaxies in five redshift bins from $z\sim0$ to $z\sim 0.5$, and the evolution of the average size as a function of redshift for different stellar mass bins. 
The size adopted is the circularized half-light radius.
They select spheroid  and disc-dominated galaxies using SED fitting and S\'ersic index.

We summarize the observational data considered and their main characteristics in table~\ref{tab:obs_data}.
For all data considered, galaxies are classified primarily using color or SFR (based on SED fitting) estimates.
Morphological selection (through the S\'ersic index) is a secondary criterion adopted only in some of the samples considered. 
Therefore, in the following analysis, the most appropriate comparison is with model SF/Q galaxies, although the selection is not necessarily the same. 

\begin{table*}
% \centerfloat
   \centering
    \caption{Summary of the observational data considered in the size--mass analysis.
    We list in different columns: the stellar mass limit, and the selection adopted to divide SF or LT from Q or ET galaxies.}
    \label{tab:obs_data}
    \begin{tabular}{ | c | c c c |  } 		\hline
	Observational data	&	Stellar mass limit & 	LT/ET selection	\\ \hline
  \citet{williams2010uds_r_z}	&	$M_*>10^{10.8}\,{\rm M_{\sun}}$	&	sSFR$=0.3/t_H$	\\
  \citet{cassata2013goods_candles_r_z}	&$M_*>10^{10}\,{\rm M_{\sun}}$	&	sSFR$< 10^{-2}\,{\rm Gyr^{-1}}$	 \\
  \citet{van_der_wel2014candles}	&$M_*>10^{9}\,{\rm M_{\sun}}$	&	U-V vs V-J	\\
  \citet{favole2018boss_decals}	&	$M_*>10^{11}\,{\rm M_{\sun}}$	&	Luminous Red Galaxies	\\
  \citet{roy2018kids}	&	$M_*>10^{10}\,{\rm M_{\sun}}$ for $z>0.2$& SED fitting + S\'ersic index \\

    \hline
	\end{tabular}
\end{table*}

\subsection{Dependence on the selection}

We divide our model galaxies in LT/ET according to their bulge over total mass ratio $B/T$, and in SF/Q according to their specific star formation rate sSFR. 
Different selections produce different results, in particular at high redshift.
Here, we analyze model predictions in detail.

\subsubsection{Size--mass relation of SF/LT galaxies}
In figure~\ref{fig:reff_ms_redshift}, both SF and LT galaxies show the same median size--mass relation. 
Their relations are in fair agreement with observations by \citet{roy2018kids} up to $z\sim0.5$ and  \citet{van_der_wel2014candles} up to redshift $z\sim2$.
Our model relation is slightly above observational estimates at low stellar masses, while at the high mass end ($M_*>10^{11}\,{\rm M_{\sun}}$) the situation is inverted.
The observed relation remains within the predicted scatter, but the model relation is somewhat flatter than observed.
As in the data, the SF/LT relation has an almost constant normalization up to redshift $z\sim1$.
At redshift $z\sim2$, it decreases, over all the stellar mass range considered, of $\sim 0.2$ dex.

\subsubsection{Size--mass relation for Q and ET galaxies}
Q and ET model galaxies show a different behavior.  
When considering Q galaxies, we obtain a relation in agreement with observations for stellar masses up to $M_*\sim10^{11.4}\,{\rm M_{\sun}}$, at all redshifts considered. 
For larger masses, the predicted relations are below the observational estimates. 
Observational data, at such large masses,  disagree (see, for example, \citealt{roy2018kids} versus \citealt{williams2010uds_r_z} at $z\sim0.5$),  but our model Q galaxies are below all of them. 
As stated above, the selections of observed samples are mainly based on sSFR (or proxies), and the agreement with observational estimates, excluded at very large masses, is encouraging. 

On the other end, ET galaxies have sizes in fair agreement with observational estimates for masses up to $M_*\sim10^{10.8}\,{\rm M_{\sun}}$, for $z<0.5$. 
As in the case of Q galaxies, at larger masses the model relation is below the observational data. 
However, in the case of ET galaxies, this under-estimation extends to lower masses for high redshift ($z>1$). 
The relations predicted for ET galaxies at $z\sim 1$ and $z\sim 2$ lie within the scatter distribution of galaxies observed by \citet{cassata2013goods_candles_r_z}.
These data are based on a sSFR selection, and use circularized radii, that could underestimate sizes with respect to the major-axis estimate used by \citet{van_der_wel2014candles}.

We further analyze ET galaxies, selecting those whose bulges were formed mainly through mergers (ET MB). This selection excludes unrealistically small bulges formed through disk instabilities. 
The result is shown as a dashed green line in figure~\ref{fig:reff_ms_redshift} (only for MRI galaxies).
ET MB galaxies follow nicely the relation found for Q galaxies, up to redshift $z\sim 2$, in good agreement with observations (with the exception of very large masses).
This result emphasizes again the influence of disk instabilities in lowering the sizes of ET galaxies, in particular in the mass range $M_*\in[10^{10};\, 10^{11}]\,{\rm M_{\sun}}$.
In the case of Q galaxies, bulges form mainly through disk instabilities, but they have a $B/T<0.5$, and their size is not influenced much by the bulge formation channel.

\subsubsection{The relation for central galaxies}
In \citet{zoldan2018size_j}, we have shown that model central and satellite ET galaxies have different size--mass relations at $z=0$, in contrast with observational evidence \citep{huertas2013size}. 
This is due to the disk instability treatment, that affects more strongly the sizes of central galaxies. 
For clarity, we have not shown the relations for only central galaxies in figure~\ref{fig:reff_ms_redshift}, but we show them in figure~\ref{fig:reff_ms_redshift_centrals}, in appendix~\ref{appendix:r_ms_centrals}.
SF/LT centrals have sizes indistinguishable from those of the entire SF/LT population, and the same is valid for Q central galaxies. 
ET central galaxies, instead, exhibit the same behavior found at redshift  0: in the mass range $M_*\in[10^{10.2}\,;10^{11}]\,{\rm M_{\sun}}$, the size--mass relation exhibits a ``dip'' corresponding to a median $\log_{10}(R_{1/2}\,[{\rm kpc}])\sim -0.3$. 

At high redshift ($z\sim2.07$), the Q population is dominated by satellite galaxies, with very few centrals, while the ET population is dominated by centrals that have grown their bulge through disk instability, and thus are very small. 
This is not surprising, as at these early times mergers are not enough to represent a relevant bulge formation channel.

%------------------------------------------------
%----------EVOLUTION OF SIZE AND NUMBER DENSITY---
%------------------------------------------------
\section{Evolution of size and number density of Q galaxies}
\label{sec:r_n_z}
%------- Reff vs z of all galaxies; ang_mom/j_integration/halo_prop/plot_r_z_mbins.py (profiles made with calc_less_for_all_MRI.py)
\begin{figure*}
    \includegraphics[trim=0.5cm 0.cm 1.5cm 1cm, clip, width = 0.99\linewidth]{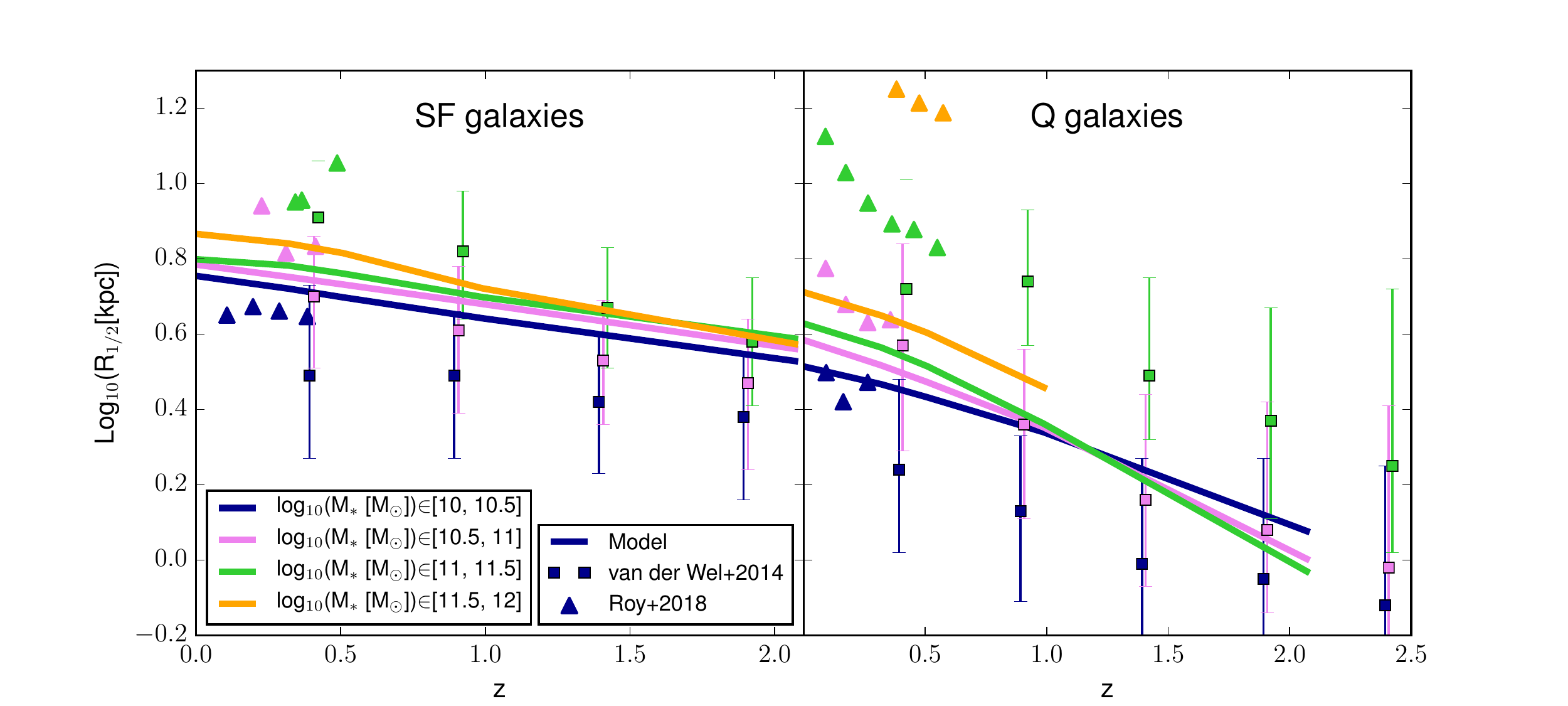}
    \caption{The evolution of the median size of galaxies in different stellar mass bins. 
    SF galaxies are shown in the left panel, while Q galaxies are shown on the right. 
    We consider four stellar mass bins, shown with different colors, as indicated in the legend. 
    Model relations are shown as solid lines, while observational estimates are shown by squares \citep{van_der_wel2014candles} and triangles \citep{roy2018kids}.
    The 16-84th percentiles of the distributions are not shown for clarity, but the scatter is $\sim 0.2$ dex over the entire redshift range shown.
    }
    \label{fig:r_z}
\end{figure*}
In figure~\ref{fig:r_z}, we show the evolution with redshift of the half-mass radius of galaxies selected in fixed stellar mass bins (different colors). 
Galaxies are classified as SF (left) and Q (right panel), according to a sSFR$=0.3/t_H$ threshold. 
We do not show results for a $B/T$ selection for clarity, as it does not match the selections adopted in the observational samples considered here.
We use, as a term of comparison, median observed circularized radii by \citet[][squares]{van_der_wel2014candles} and by \citet[][triangles]{roy2018kids}. 
The stellar mass bins used in \citet{roy2018kids} are not exactly matching those used in the figure, but are slightly shifted towards larger masses ($\sim 0.2$ dex) . 

We find a good agreement with observations for the lowest mass bins considered ($M_*<10^{11}\,{\rm M_{\sun}}$) for SF galaxies, while at larger masses model predictions tend to stay below observational measurements.
Q model galaxies reproduce quite well the observational estimates for low masses, although the lowest stellar mass bin ($M_*\in[10^{10};\,10^{10.5}]\,{\rm M_{\sun}}$) is above observational estimates at high redshift ($z>0.5$). 
For larger masses ($M_*>10^{11}\,{\rm M_{\sun}}$), model Q galaxies are much smaller than observed at all cosmic epochs.
These trends are consistent with what discussed in the previous section. 

\begin{figure}
    \includegraphics[trim=1cm 0.cm 1.5cm 0.5cm, clip, width = 0.99\linewidth]{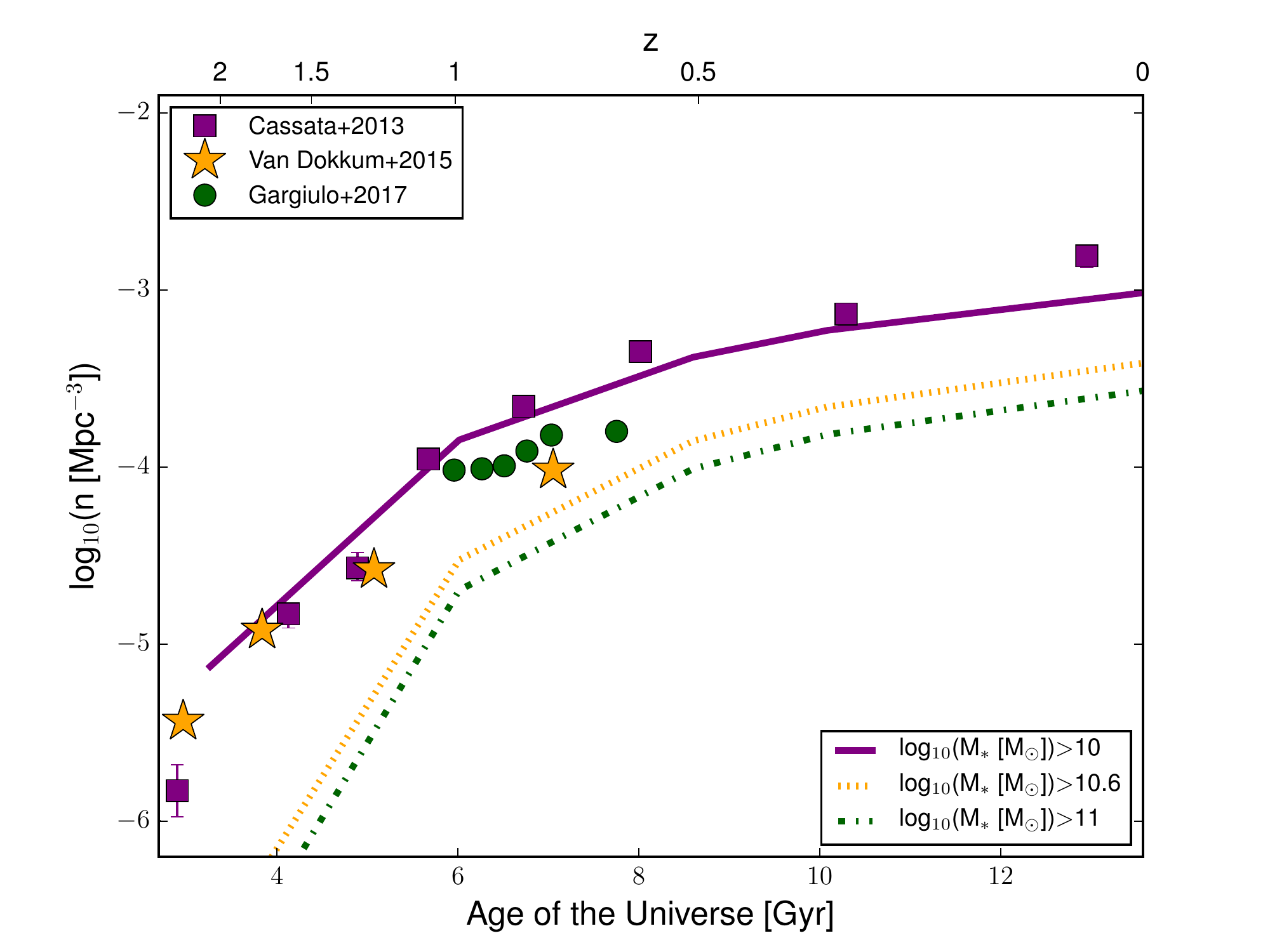}
    \includegraphics[trim=1cm 0.cm 1.5cm 0.5cm, clip, width = 0.99\linewidth]{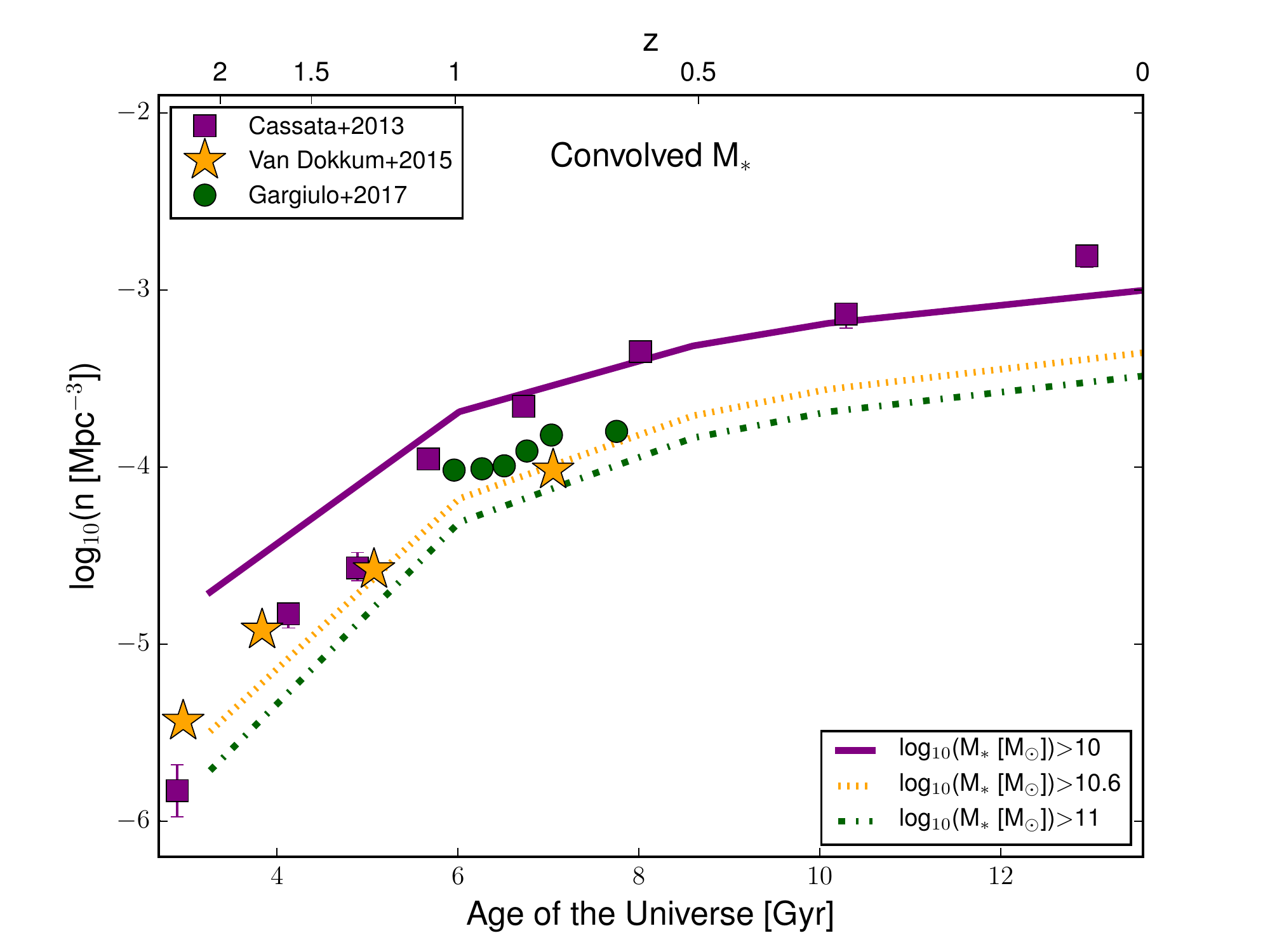}
    \caption{Co-moving number density evolution of Q model galaxies.
    Different line styles and colors represent different stellar mass cuts: $M_*>10^{10}\,{\rm M_{\sun}}$ as a dashed solid purple line; $M_*>10^{10.6}\,{\rm M_{\sun}}$ as a dot-dashed orange line; and $M_*>10^{11}\,{\rm M_{\sun}}$ as a dashed green line. 
    These stellar mass cuts correspond to those adopted in observational studies whose results are shown using symbols with the same color: \citet{cassata2013goods_candles_r_z} as purple squares,  \citet{van_dokkum2015r_z} as orange stars, \citet{gargiulo2017r_z} as green circles.
    In the top panel, we show the direct model outputs, while in the bottom panel, we show the same results but for stellar masses convolved with a log-normal error distribution, as detailed in the text.
    }
    \label{fig:num_z_ET}
\end{figure}

We consider now the evolution of the co-moving number density of Q galaxies, and show our results in the top panel of figure~\ref{fig:num_z_ET}.
To have a fair comparison with observations, we select Q galaxies adopting different stellar mass thresholds. 
Solid, dotted and dash-dotted lines show predictions corresponding to $M_*>10^{10}\,{\rm M_{\sun}}$ \citep[as in][]{cassata2013goods_candles_r_z}, $M_*>10^{10.6}\,{\rm M_{\sun}}$ \citep[as in][]{van_dokkum2015r_z}, and $M_*>10^{11}\,{\rm M_{\sun}}$ \citep[as in][]{gargiulo2017r_z}.  
%------- number density evolution; ang_mom/j_integration/halo_prop/plot_density_evolution.py (profiles made with plot_r50_things_multi.py and ../redshift/plot_r_DI....)
\citet{van_dokkum2015r_z} selected galaxies from the 3D-$HST$ project \citep{brammer20123DHST}, according to their position in the $U-V$ versus $V-J$ diagram (the considered size is the circularized half-light radius $r_e$). 
In  \citet{gargiulo2017r_z}, ET galaxies are from the VIPERS project \citep{scodeggio2018VIPERS}, with spectroscopic redshift $0.5<z<1$. 
Their galaxies are selected according to the rest-frame NUV--$r$ vs $r$--$K$ diagram.
We show observational data using different symbols, with color corresponding to each mass cut considered. 

When considering the mass thresholds adopted by \citet{cassata2013goods_candles_r_z}, our model Q number density is in agreement  with the corresponding observational estimates. 
When considering the mass thresholds adopted by \citet{van_dokkum2015r_z} and  \citet{gargiulo2017r_z}, model predictions are offset low with respect to observational estimates, at all cosmic epochs.
If we consider the uncertainty in the stellar mass measurement, adding Gaussian random error to $\log_{10}(M_*)$ with a standard deviation of $\sigma_{*}=0.25$ dex, the resulting number density evolution changes significantly, in particular at high redshift, as shown in the bottom panel of figure~\ref{fig:num_z_ET}. 
In this case, the high redshift Q galaxy population increases significantly, bringing the number density evolution at high redshift in better agreement with observational data.

\begin{figure}
    \includegraphics[trim=1cm 0.cm 1.5cm 0.5cm, clip, width = 0.99\linewidth]{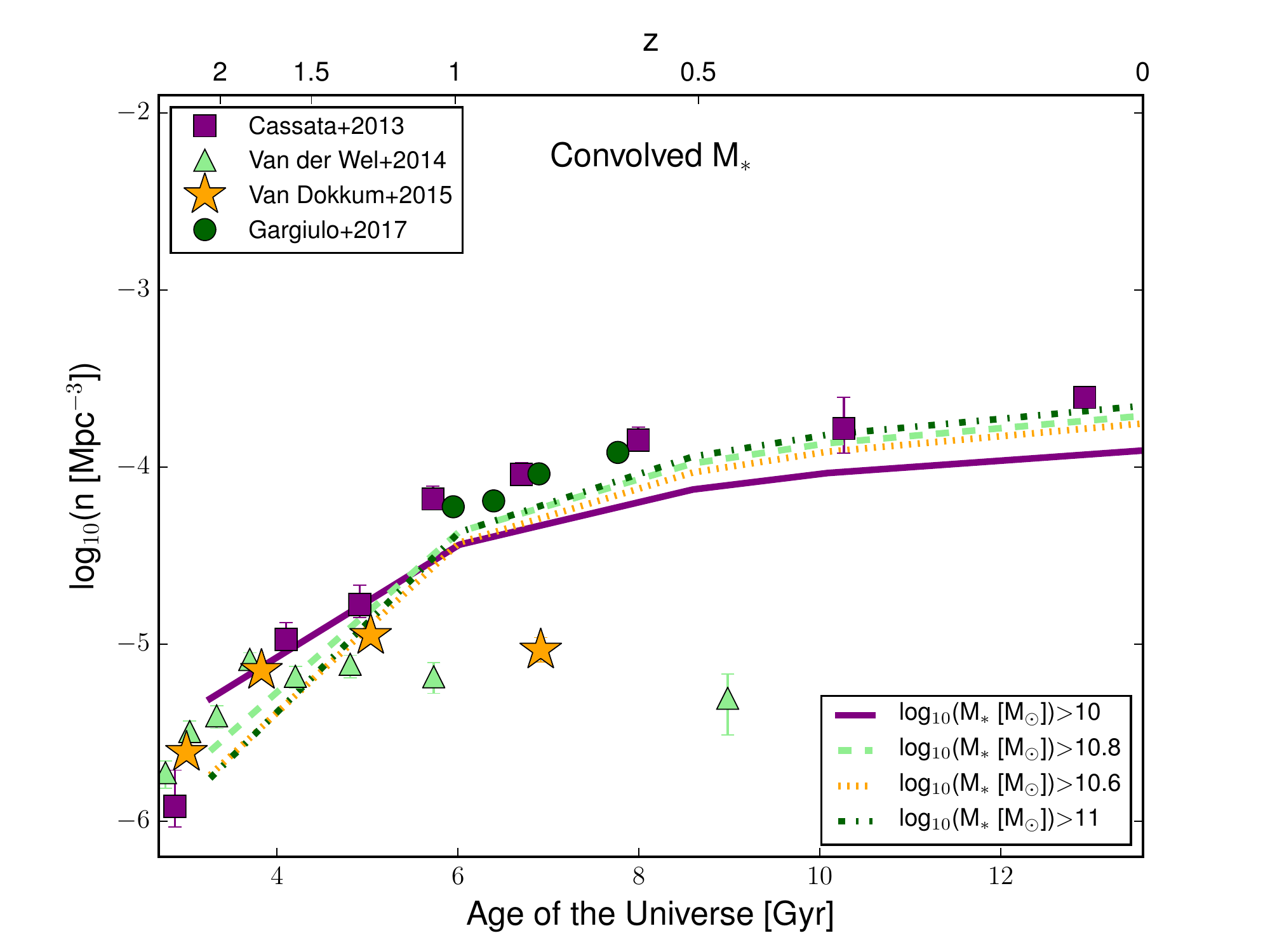}
    \caption{Same as the bottom panel of figure~\ref{fig:num_z_ET}, but for Q compact galaxies. 
    Model stellar masses have been convolved with a log-normal distribution of standard deviation 0.25 dex.
    Model compact galaxies are selected using the same methods adopted in the observational studies (different line styles, colors corresponding to observations). 
    Observational data come from the same studies considered in figure~\ref{fig:num_z_ET}, but include only  compact Q galaxies. 
    Additional observational data are by \citet{van_der_wel2014candles}, shown as light green triangles. 
    }
    \label{fig:num_z_C}
\end{figure}

In this context, it is also interesting to analyze the number density evolution of Q ``compact'' galaxies.
These are numerous at high redshift, and become a small fraction of the total population at redshift 0. 
Observationally, compact galaxies are selected following different criteria, based either on galaxy size or galaxy stellar density. 
We mimic these different selections, and plot the predicted number density evolution in figure~\ref{fig:num_z_C}. 
Colors and styles are as in figure~\ref{fig:num_z_ET}, with the addition of observational measurements (and the corresponding model selection) by \citet[light green dashed lines and triangles]{van_der_wel2014candles}.
We show the relations for model stellar masses convolved with a log-normal distribution. 
%------- number density evolution of compact galaxies; ang_mom/j_integration/halo_prop/plot_density_evolution.py (profiles made with plot_r50_things_multi.py and ../redshift/plot_r_DI....)

In \citet{cassata2013goods_candles_r_z}, the selection of compact galaxies is performed by comparing the galaxy position on the  size--mass plane to that of SDSS Q galaxies at redshift 0: those with radius below the 16th percentile of the SDSS distribution are considered compact.
The selection at higher redshift is still done compared to the SDSS at redshift 0. 
Using the same selection on model galaxies, we obtain a median number density with the same shape of the observational estimates, but the predicted relation is slightly below the data over the entire time interval considered. 
If we use a higher sSFR threshold for the Q selection (sSFR$\sim 10^{-10}\,{\rm yr^{-1}}$), we find a  better agreement for the compact number density evolution, but this affects also the total Q number density, bringing it slightly above observational estimates. 

In \citet{gargiulo2017r_z}, compact galaxies are selected according to their stellar mass density: for the purposes of this work we group together all their galaxies with stellar density larger than $>1000\,{\rm M_{\sun}\, pc^{-2}}$. 
We calculate the mean stellar surface density of model galaxies using $\Sigma_* = M_*/(2 \pi R_{1/2}^2)$, as done in the observational work. 
Again, we find the predicted trend is similar to that of the observational estimates in terms of slope, but the normalization is slightly below the data at all cosmic epochs considered. 

In \citet{van_der_wel2014candles}, compact galaxies are defined as galaxies with $R_{1/2}/(M_*/10^{11}\,[{\rm M_{\sun}}])^{0.75}< 2.5\,{\rm kpc}$. 
At large stellar masses, this selection is quite similar to that by \citet{cassata2013goods_candles_r_z}.
In \citet{van_dokkum2015r_z}, compact galaxies are selected to be massive ($\log_{10}(M_*)>10.8$) and to satisfy the relation $\log_{10}(R_{1/2})<\log_{10}(M_*) -10.7$. 
For these two last selections, we find many more compact model galaxies with respect to observational estimates for $t_H>5\,{\rm Gyr}$.
\citet{gargiulo2016compact} argue that the difference between estimates by \citet{van_der_wel2014candles} and \citet{gargiulo2017r_z} can be largely ascribed to the different definition of galaxy sizes. 
The different selections may also play a role.

\section{Specific angular momentum evolution}
\label{sec:j_z}
In a previous work, we have analyzed the stellar specific angular momentum versus mass relation for SF, Q,  LT and ET galaxies at $z=0$ \citep{zoldan2018size_j}. 
We have computed different estimates of the specific angular momentum of model galaxies, to account for different sources of uncertainties, like projection, inclination and integration techniques. 
We have found a good agreement with available observational data. 
Here, we extend this analysis to higher redshifts, referring to the original paper for more details on the technical aspects of this calculation \citep{zoldan2018size_j}. 
%------- j* vs M* with j* from the profiles, at various redshifts; ang_mom/j_integration/plot_test_redshift_multi.py (profiles made with calc_more_jaffe.py)
\begin{figure*}
    \includegraphics[trim=2cm 0.8cm 2.5cm 2cm, clip, width = 0.99\linewidth]{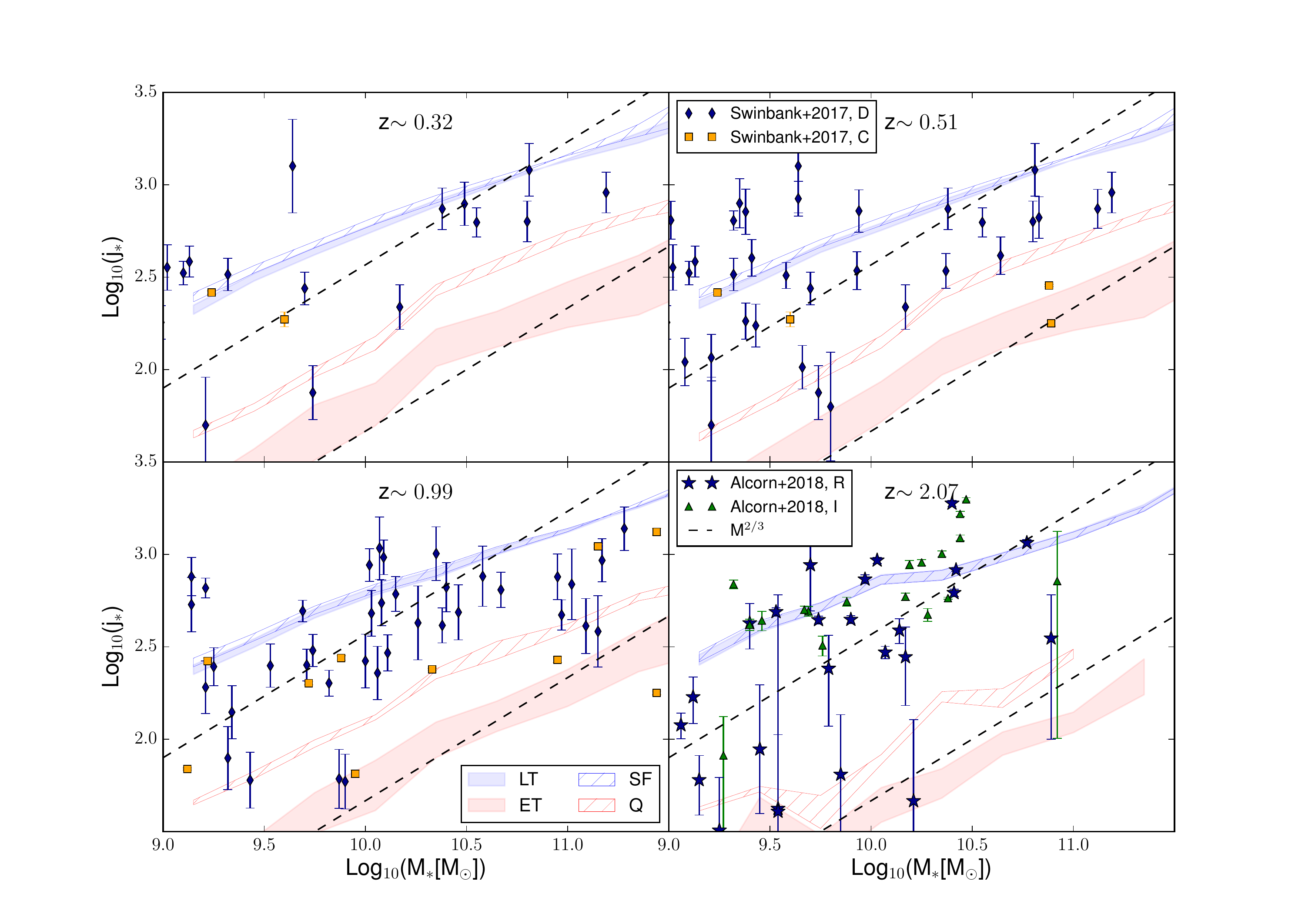}
    \caption{The $j_*$--$M_*$ relation for model galaxies.
      Predictions for LT and ET galaxies are shown as shaded areas (blue and red). 
      Those for SF and Q galaxies are shown as hatched areas (blue and red). 
      Different panels show different redshifts, as indicated in the legends. 
      Symbols correspond to observational measurements by \citet{swinbank2017js_z} and \citet{alcorn2018j_z} for disk dominated galaxies (D, blue diamonds), composite dynamics (C, orange squares), regular (R, blue stars) and irregular (I, green triangles). 
      Observations are selected to be close to the redshift of each panel, with a $\Delta z\lesssim 0.2$.
      The data by \citet{alcorn2018j_z} are considered to be all at $z\sim2$.
      The dashed lines represent the theoretical expectation for the slope of the $j_*-M_*$ relation, with arbitrary normalizations.}
    \label{fig:jo_ms_redshift}
\end{figure*}

In figure~\ref{fig:jo_ms_redshift}, we show the median $j_*$--$M_*$ relation for model galaxies. 
Each panel shows model results at a given redshift, as indicated in the caption. 
The median relation is represented as a shaded/dashed area, that accounts for different calculation methods and inclinations. 
Galaxies are divided in SF, LT (blue), Q and ET (red) using a sSFR$=0.3/t_H$ (SF/Q, dashed areas) or a $B/T=0.5$ (LT/ET, shaded areas) threshold. 
We do not show the scatter of model predictions, but it amounts to $\sim 0.4$ dex, independently of stellar mass.
We show observational data by \citet{swinbank2017js_z} and by \citet{alcorn2018j_z}, with symbols and colors depending on galaxy dynamical state.
Error bars show the uncertainties in the measurements. 
\citet{swinbank2017js_z} measured the dynamics of 400 SF galaxies at redshift $z\sim 0.28-1.65$, using MUSE and KMOS integral field spectrographs. 
They measured the [OII] and H$\alpha$ emission lines to infer the velocity and dispersion of the stars in each galaxy. 
They classified galaxies according to their internal dynamics and importance of rotational velocity with respect to dispersion. 
Galaxies are considered {\it rotationally supported (D, disks)} if the dynamics appears regular (blue diamonds); {\it irregular (I)}, if they have a complex velocity field and morphology; {\it unresolved (U)} or {\it composite/major merger (C)}, if they are made of two or more interacting galaxies (orange squares).
\citet{alcorn2018j_z} selected a sample of galaxies at $z=1.7-2.5$ from the ZFIRE survey \citep{nanayakkara2016zfire}, identifying SF galaxies  using their UVJ colors. 
Using CANDELS data, processed by the 3D-$HST$ team \citep{skelton20143DHST}, they evaluated the sizes of galaxies. 
The rotational velocity was calculated from the H$\alpha$ emission line, for {\it regular (R,} blues stars) and {\it irregular (I,} green triangles) galaxies. 
In both studies, the galaxy stellar specific angular momentum is evaluated using the approximated formula proposed in \citet{romanowsky2012js}.
This depends on the rotational velocity at 2 half light radii, a parameter that depends on the S\'ersic index, the half light radius, and a de-projection factor. 
In each panel of figure~\ref{fig:jo_ms_redshift}, observational data are selected to be at a maximum redshift distance of $\Delta z\sim0.2$ from that shown in the caption, while data by \citet{alcorn2018j_z} are considered to be all at $z=2$.
We show,  as a reference, two dashed lines representing the theoretical expectation for the slope of the relation \citep[][the normalization is arbitrary]{romanowsky2012js}. 

The relations for SF and LT galaxies show little to no evolution with redshift.
In all panels, they lie within the observational scatter. 
A certain number of observed galaxies are below our model relation, and this is more evident at low masses at high redshift. 
These galaxies extend down to the relation expected for Q galaxies. 
In our model, we find that at $z\sim2$ from $1$ to $10\%$ of the LT/SF galaxies  with $M_*<10^{10}\,{\rm M_{\sun}}$ have $\log_{10}(j_*)<2$.
The observational error bars are large, and observational measurements are typically indirect, as they are based on a formula calibrated at $z=0$, while our model estimates include also a direct integration of $j_*$.

Q/ET galaxies are characterized on average by a decreasing $j_*$ with increasing redshift, a trend more clear in the case of ET galaxies. 
Q galaxies have a higher normalization than that obtained for ET galaxies, at all stellar masses and at all redshifts. 
This reflects the results discussed for the size-mass relation.

%------------------------------------------------
%-----------RELATION WITH DM HALO PROPERTIES-----
%------------------------------------------------
\section{Relation with DM halo properties}
\label{sec:r_halo_z}
In previous work, we have found a correlation between halo spin parameter and cold gas content \citep{zoldan2017MNRAS.465.2236Z}, and between the cold gas content and galaxy specific angular momentum \citep{zoldan2018size_j}. 
In this section we analyze the dependence of galaxy sizes on halo properties, such as radius and spin parameter, and we compare our findings with observational measurements.

\subsection{Dependence on halo $R_{200}$}
\citet{huang2017r50_r200} analyzed data from the CANDELS survey \citep{galametz2013CANDLES,guo2013CANDELS}, and selected the 20$\%$ galaxies with the highest and lowest estimates of sSFR and S\'ersic index. 
They evaluated $M_{200}$ corresponding to the parent halo of each galaxy using abundance matching, and evaluated the halo radius using equation~\ref{eq:r200}. 
They find that SF/LT and Q/ET galaxies lie on almost parallel relations in the $R_{1/2}$--$R_{200}$ plane, with Q/ET galaxies having a $\sim2$ times smaller $R_{1/2}$, independently of redshift. 
Results do not vary significantly assuming different Stellar Mass--Halo Mass (SMHM) relations.
The slope of these relations is similar to that found by \citet{kravtsov2013r50_r200}, who analyzed hundreds of nearby galaxies with a wide range of morphologies, masses and environments, using a similar approach. 

\begin{figure*}
    \includegraphics[trim=2cm 1cm 2.5cm 2cm, clip, width = 0.99\linewidth]{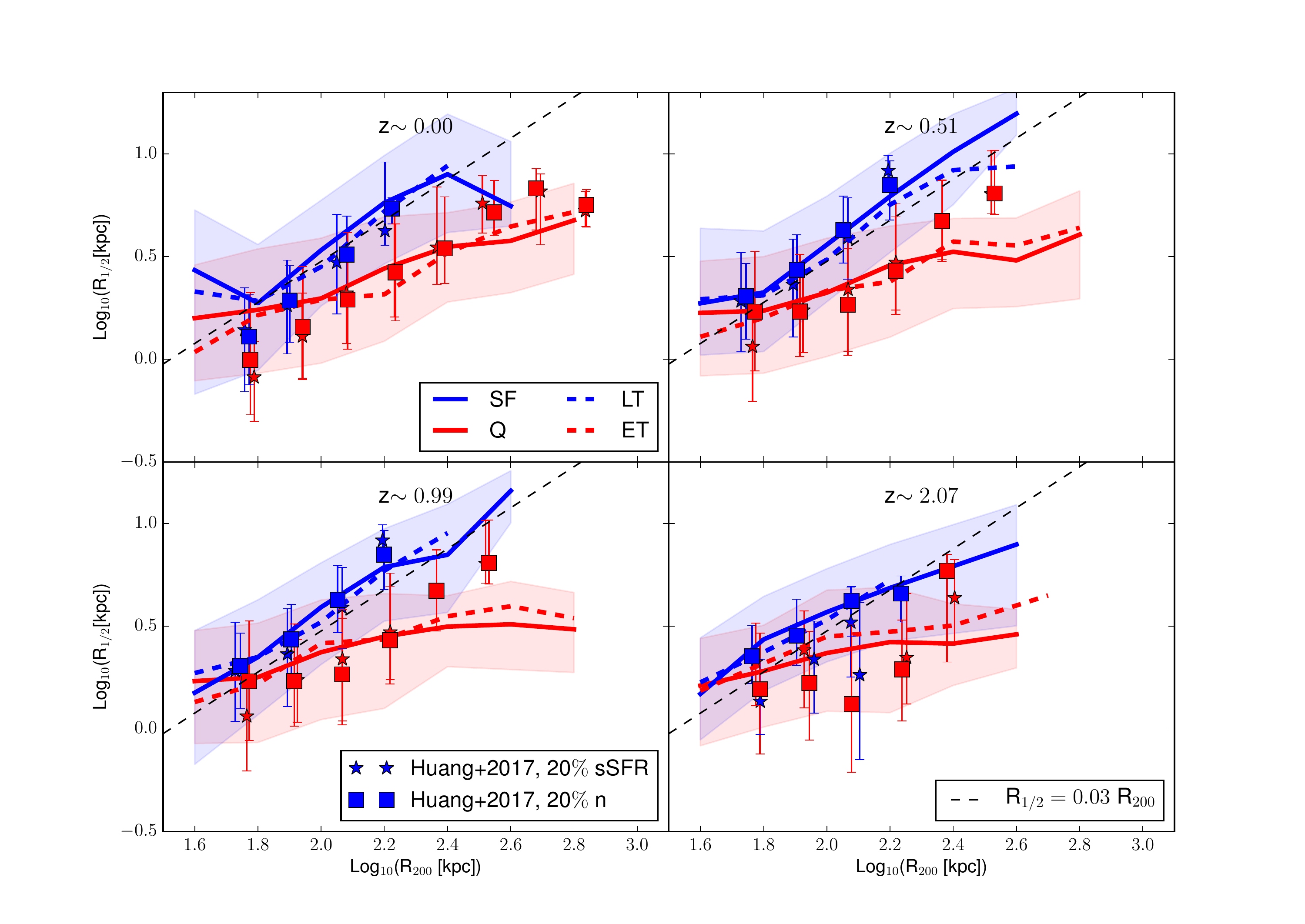}
    \caption{The  $R_{1/2}$--$R_{200}$ relation at various redshifts (different panels), for galaxies selected to be in the highest and lowest 20\% of the distributions in sSFR (solid lines) and in $B/T$ (dashed lines).
    Symbols with error-bars are  observational estimates by \citet[][model 1]{huang2017r50_r200}. Different colors correspond to the highest and lowest 20\% of the measured  sSFR (stars) and S\'ersic indexes (squares).
    As a reference, we show the relation $R_{1/2}=0.03R_{200}$ as a dashed black line. }
    \label{fig:r200_r50_z}
\end{figure*}

In figure~\ref{fig:r200_r50_z}, we show the $R_{1/2}$--$R_{200}$ relation at four redshifts.
We select SF/Q and LT/ET galaxies (blue/red) considering the 20\% tails of the distributions in sSFR (solid lines) and $B/T$ (dashed lines).
To have a compatible number of galaxies picked from the MRI and MRII (galaxies with $M_*<10^{10}\,{\rm M_{\sun}}$ are selected from the MRII) we consider a sub-volume of the MRI of the same size of the MRII. 
In this figure, we include satellite galaxies, whose $R_{200}$ is that of the halo  at the last time its was identified.
Results obtained using only central galaxies are similar.
For each redshift, we show the estimated relations by \citet[][model 1]{huang2017r50_r200}, for both selections based on  S\'ersic index (squares) and  on sSFR (stars).
%------- Reff vs Rvir of LT/ET galaxies at different redshitfs; ang_mom/j_integration/halo_prop/plot_r50_things_multi.py (profiles made with calc_more_jaffe.py)

Model SF and LT galaxies show a $R_{1/2}$--$R_{200}$ in good agreement with  observational estimates. 
We find that $\sim 68 \%$ of SF galaxies have  $\log_{10}(R_{200}\, [{\rm kpc}])$ between $1.99$ and $2.14$, while  $\sim 68 \%$ of Q galaxies  lie in the range $\log_{10}(R_{200}\, [{\rm kpc}])\in[1.84; 2.11]$ at redshift 0.
At higher redshift these ranges shrink. 

Q and ET galaxies are in good agreement with observational measurements for $\log_{10}(R_{200}\, [{\rm kpc}])<2.5$, while at larger halo radii they lie slightly below the observed relation. 
At redshift 0, model galaxies tend to be slightly larger than observed for small values of $R_{200}$, but still within the measured scatter. 
At redshift $z\sim2$, the median relations  of model Q and ET galaxies are above the observational measurements for $2<\log_{10}(R_{200}\, [{\rm kpc}])<2.3$.
Observational data, in this range, present a small dip we do not find in our model, maybe due to statistical fluctuations.
At redshift $z\sim0$, about $\sim 68 \%$ of the Q population lies in the range $\log_{10}(R_{200}\, [{\rm kpc}])\in[1.44; 2.17]$, and  about $\sim 68 \%$ of the ET population lies in the range $\log_{10}(R_{200}\, [{\rm kpc}])\in[1.55;2.31]$. 
In the case of ET galaxies, the predicted median relation does not vary significantly when selecting bulges formed mainly through mergers, i.e. it does not depend on the bulge formation channel.

\subsection{Dependence on halo $R_{200}\lambda$}

\begin{figure*}
    \includegraphics[trim=2cm 1cm 2.5cm 2cm, clip, width = 0.99\linewidth]{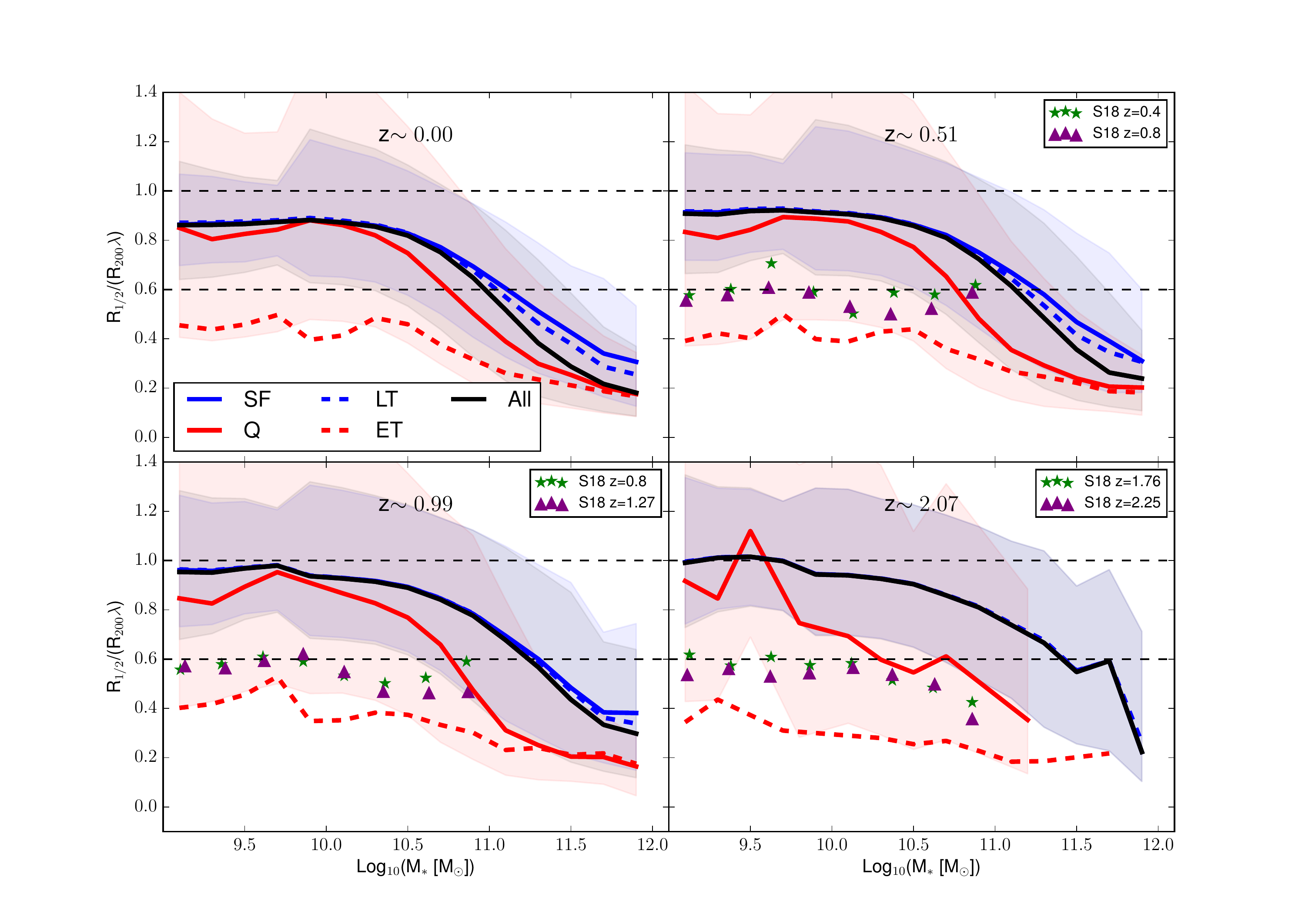}
    \caption{The  $R_{1/2}/(R_{200} \lambda)$--$M_*$ relation at various redshifts (different panels), for galaxies  selected according to their sSFR (SF/Q, solid lines) and their $B/T$ (LT/ET, dashed lines). 
    SF and LT galaxies are shown in blue, while Q and ET galaxies are shown in red.
    The median for the entire population is shown as a black solid line. 
    Shaded areas represent the 16-84th percentiles of the distributions.
    Two horizontal dashed lines are shown at 1 and 0.6 to guide the eye.
    Observational data by \citet{somerville2018r50_r200} are shown for a qualitative comparison as different colored symbols, according to their redshift, as indicated in the legend. 
    }
    \label{fig:r50r200L_Ms_z}
\end{figure*}

As explained in the introduction, if we assume that baryons are initially coupled to the DM halo dynamics, the final $R_{1/2}$ is proportional to $R_{200} \lambda$.
In the previous section, we have seen that $R_{1/2}$ is proportional to $R_{200}$, with the slope depending on the galaxy type (SF/LT or Q/ET).
In figure~\ref{fig:r50r200L_Ms_z}, we show the ratio $R_{1/2}/(R_{200} \lambda)$ versus $M_*$ for different redshifts (different panels). 
The median relations are shown as lines, solid for a SF/Q (blue/red) selection (based on sSFR$=0.3/t_H$), and dashed for a LT/ET (blue/red) selection (based on a $B/T =0.5$ threshold). 
We show also the median relation for the entire population as a solid black line. 
The shaded areas represent the 16-84th percentiles of the distributions. 
We show two horizontal lines at 1 and 0.6 to guide the eye.

%------- Reff/(R200*lambda) of LT/ET galaxies at different redshitfs; ang_mom/j_integration/halo_prop/plot_r50_things_multi.py (profiles made with calc_more_jaffe.py)

For our model, we have considered $R_{200}$ and $\lambda$ evaluated at the last time each galaxy was a central galaxy.
This was done because after accretion, satellites subhalos are efficiently disrupted by tidal interactions with their parent halo, and their spin is difficult to evaluate in case the number of particles is small. 
Furthermore, observational studies infer halo sizes from abundance matching techniques without differentiating central and satellite galaxies. 
SF and LT galaxies have a $R_{1/2}/(R_{200} \lambda)$ almost independent on stellar mass up to $M_*\sim 10^{10.5}\,{\rm M_{\sun}}$, and then slowly decreasing at larger masses. 
At low and intermediate masses $R_{1/2}/(R_{200} \lambda)\sim 0.8$ ($\sim1$ at redshift $\gtrsim1$): halo and stellar dynamics, in these galaxies, are not far from the classical theoretical expectation.
At $z=0$, low mass ($M_*<10^{10}\,{\rm M_{\sun}}$) Q and ET galaxies behave differently: 
the former are similar to SF/LT galaxies, retaining $80\%$ of the halo dynamics, while for the latter the retention factor is only $\sim 20\%$. 
At large masses, the Q and ET relations become similar, with a $R_{1/2}/(R_{200} \lambda)\sim 0.2$.
When considering only central galaxies, the difference between Q and ET galaxies is reduced, with Q galaxies having an average $R_{1/2}/(R_{200} \lambda)\sim 0.5$ at low masses.
This behavior is similar at higher redshift. 
The relation representing the entire population reflects the prevalence of LT/SF or Q/ET galaxies in the stellar mass range considered: at low stellar masses SF/LT galaxies represent the majority of the galaxy population, while at larger stellar masses Q/ET galaxies dominate.

In figure~\ref{fig:r50r200L_Ms_z}, we show also the observational estimates by \citet{somerville2018r50_r200}, using different symbols and colors corresponding to different redshift bins. 
The comparison with these observational measurements can only be qualitative in this case, because halo radii are computed adopting different definitions: assuming a different cosmology, and using a different halo finder. 
These authors analyzed both the $R_{1/2}/R_{\rm vir}$--$M_*$ and the $R_{1/2}/(R_{\rm vir} \lambda)$--$M_*$ relations obtained for galaxies from GAMA  at redshift $z\sim0.1$ \citep{liske2015GAMA_DR2}  and CANDELS at higher redshifts \citep{grogin2011CANDELS,koekemoer2011candles}. 
They evaluated the 3D stellar half mass radius ($R_{1/2}^{3D}$), accounting for the  effects of triaxiality and projection. 
The estimates shown by \citet{somerville2018r50_r200} adopt the radius $R_{\rm vir}$ that encloses a spherical overdensity equal to $\Delta_{\rm vir} = 18 \pi^2 + 82x - 39x^2$ times the critical density, where $x=\Omega_m(z) -1$, and $\Omega_m(z)$ is the matter density relative to the critical density at redshift $z$ \citep{bryan1998Dvir} ($R_{\rm vir}$ is systematically larger than $R_{200}$ adopted in our study). 
$R_{1/2}^{3D}$ and $R_{\rm vir}$ are then linked using a SHAM approach, but with a ``forward'' modeling, opposite to the ``backward'' modeling used by \citet{kravtsov2013r50_r200} and \citet{huang2017r50_r200}. 
The ``forward model'' consists in populating the halos of a DM simulation with galaxies, using a specific  stellar mass--halo mass relation.
For a fixed stellar mass bin, the median values of $<R_{\rm vir}>$ and $<R_{\rm vir}\lambda>$ are evaluated from all simulated halos containing galaxies in the bin under consideration. 
The median $<R_{1/2}^{3D}>$ in the same fixed stellar mass bin is evaluated from the galaxies in the observational sample. 
% In contrast, the ``backward'' modeling assigns a halo mass to the stellar mass of each observed galaxy, using a stellar--halo mass relation. 
% Then, the medians are evaluated directly on the $R_{1/2}/R_{\rm vir}$ or $R_{1/2}/(\lambda R_{\rm vir})$ of each galaxy, in a fixed stellar mass bin.}

\citet{somerville2018r50_r200} find that $R_{1/2}/R_{\rm vir}$  and $R_{1/2}/(R_{\rm vir} \lambda)$ are almost independent of mass at $z=0$, but not at high redshift, where massive galaxies have lower values than less massive galaxies. 
As their radius is systematically larger than our $R_{200}$, we expect our model predictions to be systematically larger than their observational estimates. 
In fact, they find $R_{1/2}^{3D}/(R_{\rm vir} \lambda)$ between $0.5$ and $0.8$. 
For our model galaxies, the median value of $R_{1/2}/(R_{\rm vir} \lambda)$ at low masses grows with increasing redshift, and  decreases with increasing stellar mass. 
Observational estimates show this trend only at high redshift.

%------------------------------------------------
%------------------DISCUSSION-------------------
%------------------------------------------------
\section{Discussion}
\label{sec:discussion}
In this work, we have used the state-of-the-art semi-analytic model  described in 
\citet{xie2017sam} and \citet{zoldan2018size_j} to investigate the evolution of the half-mass radii and specific angular momenta of galaxies, and their dependence on galaxy and halo properties. 
We find a good agreement with observational measurements for SF/LT galaxies.
In the case of Q/ET galaxies, several issues are highlighted by the comparison with observations.
We find that the selection method affects the median size of the Q/ET population, and that at large stellar masses the predicted sizes are smaller than observed.
At the same time, model Q compact galaxies exhibit a number density evolution similar to that inferred in observational studies.
We discuss our findings in more detail in the following subsections.

\subsection{Size and selection}
We have analyzed the size--mass relation evolution for SF, Q, LT and ET galaxies.
The relations for SF and LT galaxies are in fair agreement with observations at all redshifts considered: their slope does not evolve significantly, and their normalization is almost constant up to $z\sim2$, when it starts to decrease slightly. 
These results support classical theoretical models predicting $R_{\rm disk}\propto \lambda R_{200}$.

When considering Q and ET galaxies, we find that the adopted selection has a strong impact on the resulting size--mass relation.
Q galaxies are in quite good agreement with observations up to high redshift, for $M_*<10^{11}\,{\rm M_{\sun}}$. 
ET galaxies are offset low with respect to the observational estimates at all redshifts considered.
This under-prediction at intermediate masses is driven by the predominance of bulges formed mainly through disk instabilities. 
At early stages of galaxy evolution, mergers are still not the main channel for bulge formation, and the galaxies selected by their $B/T$ are thus mainly formed through disk instabilities. 
These bulges in central galaxies (that outnumber satellite galaxies at high redshift) are unrealistically small, as shown in previous work focused on $z=0$ \citep{zoldan2018size_j}.
Considered that observational studies adopt selection criteria closer to a sSFR selection (color, SED fitting, direct sSFR estimates), we find a remarkably good agreement with our model predictions for SF and Q galaxies.

\subsection{Size of Q and ET galaxies at high masses}
\label{sec:size_high_mass}
At large masses, our predicted galaxy sizes for Q and ET galaxies are offset significantly below observational estimates. 
In the case of ET galaxies, this does not depend on the bulge formation channel, as, at these high masses, bulges are mainly formed through mergers. 
Similar   findings have been obtained for independent semi-analytic models: \citet{irodotou2018sam} and \citet{lagos2018sam} show a similar flattening of the size--mass relation for ET galaxies with masses $M_*>10^{11}\,{\rm M_{\sun}}$, using models with very different prescriptions for various physical processes.

%------- size--mass relation of SF-LT/Q-ET galaxies at z=0, for the high mass end assuming different prescriptions for some selected physical processes; ang_mom/j_integration/halo_prop/plot_high_mass_models.py (profiles made with calc_more_jaffe.py)
\begin{figure*}
    \includegraphics[trim=2cm 0cm 2.5cm 1cm, clip, width = 0.99\linewidth]{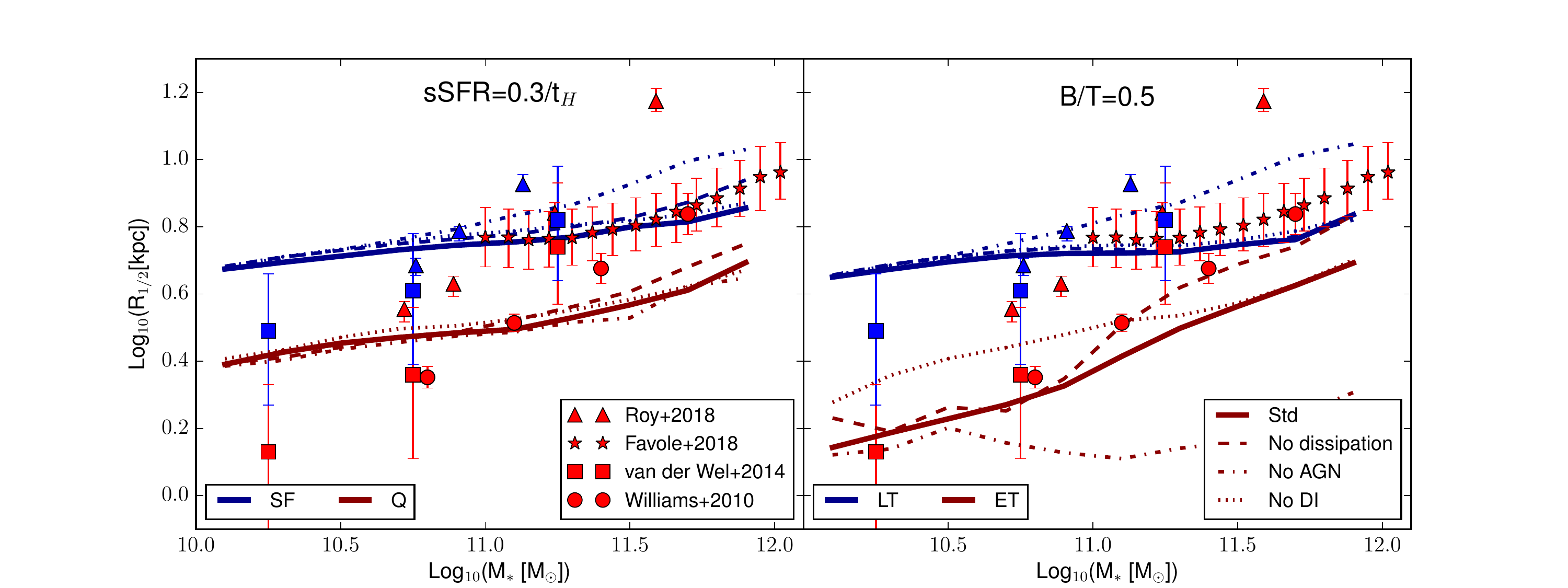}
    \caption{The  $R_{1/2}$--$M_*$ median relation at redshift $z\sim0.5$ for galaxies obtained using four model variants: our fiducial run (solid), a run without dissipation during mergers (dashed), a run without AGN feedback (dot-dashed) and one without disk instabilities (dotted lines). 
    On the left panel, we show results for SF (blue) and Q (red) galaxies, while on the right panel we show LT (blue) and ET (red) galaxies.
    We show, as a reference, observational data  corresponding to this redshift bin, as described in section~\ref{sec:obs_reff}
    }
    \label{fig:reff_ms_models}
\end{figure*}
We further investigate this issue by studying the effect of several prescriptions on the size--mass relation of model SF, Q, LT and ET galaxies. 
Some of these variants do not produce appreciable differences. 
We briefly describe them in the following. 

The first model variant considered assumes a cold gas specific angular momentum 3 times larger than that of the hot gas, for gas cooled through cold accretion. 
This is motivated by recent numerical studies \citep{stewart2011cold_accr,pichon2011cold_accr,danovich2015cold_accr}.
As in previous work \citep{zoldan2018size_j}, we find that this modification is important only for low-mass galaxies and does not influence significantly the size of most massive galaxies.

As mergers are the major channel for bulge formation at the massive end, we analyze the effect of  changing the merging time of orphan satellite galaxies by a factor $0.5$ with respect to the standard model. 
In our model, satellite galaxies survive the disruption of their host subhalo: a merging time is assigned to each orphan satellite, according to the classical dynamical friction formula \citep{binney_tremaine_2008}. 
This time is corrected using a factor $f_{\rm mer}=2$ that accounts for results from numerical simulations \citep{boylan_kolchin2008tdyn,jiang2008tdyn}. 
We tested the effect of shorter ($f_{\rm mer}=1$) or longer ($f_{\rm mer}=3$) merger times, finding very little differences in the size--mass relation, of the order of less than 0.1 dex in $\log_{10}(R_{1/2}\,[{\rm kpc}])$. 

Other model variants affect significantly the predicted size--mass relation. 
We show them in figure~\ref{fig:reff_ms_models}, for redshift $z\sim0.5$, but a similar behavior is obtained at lower and higher redshifts. 
In the left panel, we show the median size--mass relations obtained for SF (blue) and Q (red) galaxies,  while in the right panel we show LT (blue) and ET (red) galaxies. 
Different line styles represent different model variants, as indicated in the legend. 
Observational data are shown as symbols, as detailed in the legend. 
We do not attempt to re-tune model free parameters when turning off specific physical processes, because the aim of this analysis is to understand the effect of the individual prescription on the final size, and  not to reproduce the data. 

Figure~\ref{fig:reff_ms_models} clearly shows that considering dissipation during major mergers affects considerably bulge sizes.
% A prescription acting during mergers, that affects directly bulge sizes, is that regulating dissipation during major mergers.
Albeit this treatment was introduced to obtain realistic bulge sizes for small galaxies, we find that it affects also the sizes of galaxies at the high mass end. 
In figure~\ref{fig:reff_ms_models}, we show model predictions in a run without dissipation as dashed lines. 
The increase of galaxy size is modest when considering Q galaxies, and is more evident at the  massive end. 
This model variant returns larger sizes for ET galaxies, all over the stellar mass range considered. 
The increase is, however, still not sufficient to match the observational estimates, for both Q and ET galaxies. 
These results indicate that the merger treatment is likely not the  unique responsible for the under-estimation of galaxy sizes at the massive end. 

We also analyze the effect of AGN feedback, that suppresses cooling in massive halos, preventing cold gas accretion and consequent star formation at late times. 
We turn off the AGN feedback, and show the resulting relations as dash-dotted lines in figure~\ref{fig:reff_ms_models}. 
For ET  galaxies, median galaxy sizes become smaller than in our reference model, while sizes remain unchanged in the case of Q galaxies. 
For both SF and LT galaxies, switching off AGN feedback leads to larger sizes. 
It should be noted that the SF/Q and LT/ET populations are dramatically affected by the lack of AGN feedback. 
In particular, Q galaxies in the original model become SF in the version without AGN, due to the fueling of fresh new cold gas, whose cooling was suppressed in the original case. 
This also leads, at least in a fraction of these galaxies, to a regrowth of the galactic disks.
Results from hydrodynamical simulations, that include sub-grid treatments for winds driven by accretion onto super massive black holes, show a scenario that is partially in contrast with the results just discussed. 
Several studies \citep[see for example ][]{dubois2013BH_r,choi2018BH_r} have demonstrated  that AGN feedback lowers the stellar surface density and increases the size of passive galaxies, by suppressing cooling and star formation only in the very center of the galaxy, a process referred to as adiabatic expansion. 
Our model does not include an explicit prescription for this mechanical feedback, and for local central quenching. 

Finally, if we turn off the disk instability prescription (dotted lines in figure~\ref{fig:reff_ms_models}), we obtain small differences  only for masses lower than $10^{11}\,{\rm M_{\sun}}$ in the case of $B/T>0.5$.
This is not surprising, as bulges of massive galaxies ($M_*>10^{11}\,{\rm M_{\sun}}$) are formed mainly through mergers, and are thus not influenced significantly by disk instabilities. 

In conclusion, of all the prescriptions analyzed above, those affecting significantly galaxy sizes at the most massive end are dissipation during mergers and AGN (quasar mode) feedback. 
Further investigations, focused on a better implementation of these two physical processes are needed to improve our model predictions for most massive galaxies.

\subsection{Compact Q galaxies}
Several observations highlighted the existence of a large population of compact or ultra-compact Q galaxies at high redshift. 
Their existence and number density at low redshift is still matter of debate, with some studies predicting an almost constant number density of compact galaxies as a function of redshift \citep{poggianti2013a}, and others claiming that compact galaxies are very rare in the Local Universe \citep{tortora2016compact}.
Several hypotheses on the fate of these galaxies have been proposed, including an increase of their sizes due mainly to mergers (in particular multiple minor mergers) or a re-ignition of star formation due to the accretion of new gaseous material.
The definition of compact galaxies varies among different studies. 
In section~\ref{sec:r_n_z}, we showed that the number density of compact Q galaxies predicted by our model is in quite good agreement with measurements by  \citet{cassata2013goods_candles_r_z} and by \citet{gargiulo2017r_z}. 
On the other end, our model predictions over-estimate measurements by \citet{van_der_wel2014candles} and \citet{van_dokkum2015r_z}, in particular at $z<1$.

We can use our model to study the formation and evolution of compact Q galaxies, adopting the compact selection by \citet{cassata2013goods_candles_r_z}. 
First, we analyze the evolution of compact galaxies selected at high redshift.
For this purpose, we identify galaxies that are quiescent and compact at redshift $z\sim1$, and select those that survive  to redshift 0. 
We find that, on average, these galaxies formed in halos with an initial mass and specific angular momentum lower than the population with a non-compact size. 
These halos acquire most of their mass and $j_h$ very early, and then do not evolve significantly down to present.
The bulges of these galaxies are formed mainly through disk instabilities, but account for around $\sim20\%$ of the total stellar mass at $z\sim1$. 
$B/T$ grows weakly with time, remaining, on average, below $B/T\sim0.4$.
We also select Q compact galaxies at $z\sim1$, excluding the constraint of surviving down to $z=0$.  
These galaxies form in halos with low $j_h$, and consequently have a low $j_{\rm *, disk}$. 
The contribution from disk instabilities to bulge growth is very low and, at $z\sim1$, the bulge accounts for less than $\sim30\%$ of the total stellar mass. 
Around $\sim 50\%$ of these galaxies are merged into larger systems by redshift  $z\sim0.5$. 
We find similar results turning off the disk instability treatment in the model, thus these results do not depend on the particular treatment adopted for disk instabilities and on its influence on bulge sizes. 

The picture for compact galaxies evolution described above is in line with the interpretation by \citet{cassata2013goods_candles_r_z} and \citet{gargiulo2017r_z}. 
These authors conclude that puffing-up of compact sizes cannot explain alone the size distribution of the present-day passive population, and that compact galaxies cannot account for all the progenitors of present day passive galaxies.
Our model highlights that compact galaxies can remain compact down to redshift 0, or can be accreted into larger systems through mergers. 
Thus, a large fraction of today passive galaxies must be of recent formation, likely from quenching of large, star forming galaxies. 
We check explicitly the origin of $z=0$ quiescent galaxies, by following their star formation history back in time. 
We find that the median population becomes quiescent around $t_H\sim 11-12\,{\rm Gyr}$ for $M_*>10^{11}\,{\rm M_{\sun}}$, thus very recently. 

In conclusion, our model compact galaxies had the bulk of their star formation at early times, and then preserved the dynamical properties down to present. 
This applies for compact galaxies selected at whatever redshift.
Those that are compact at redshifts higher than $z>1$ have a bulge formed mainly through disk instabilities, but their bulge is small, and their small size is determined mainly by their small disk.

\subsection{The evolution of the specific angular momentum}
Currently, specific angular momenta have been measured  at redshift $z>0$ only for SF galaxies. 
The estimates are based on the empirical formula by \citet{romanowsky2012js}.
Our model SF galaxies are in good agreement with observational estimates.

A few independent theoretical studies on this subject have been carried out using hydrodynamical simulations.
\citet{lagos2017eagle_j} studied the evolution of the  $j_*(R_{1/2})$--$M_*$ relation using the EAGLE simulations \citep{schaye2015eagle}.
They find that the median relation does not evolve dramatically for low and intermediate masses. 
Their model predicts a flattening of the relation for stellar masses $>10^{10.5}\,{\rm M_{\sun}}$ at redshift 0, and for $M_*>10^{10.3}\,{\rm M_{\sun}}$, at higher redshifts. 
This flattening is caused by the predominance, at these masses, of dispersion dominated galaxies.
In our model, this corresponds to the relation found for ET galaxies, that has indeed a  considerably lower normalization with respect to that predicted for LT galaxies.
\citet{lagos2017eagle_j} also find that the final specific angular momentum of a galaxy of given stellar mass does not depend strongly on the number of mergers with merger mass ratio larger than $0.1$.
We find similar results for our model galaxies, with $j_*$ decreasing only slightly with number of mergers (this is mainly due to the bulge growth during mergers). 

\citet{teklu2015magneticum_j} analyzed the specific angular momentum of stars in the Magneticum simulation \citep{hirschmann2014magneticum}. 
They have shown that cold gas and DM have proportional specific angular momenta, with a scatter independent of stellar morphology or redshift. 
Stars also have angular momentum proportional to that of the cold gas, but the proportionality factor depends, in this case, on the morphological component analyzed: disks have $j_*$ similar to that of cold gas, while spheroids have lower angular momentum than cold gas. 
This difference remains almost constant with redshift. 
These findings are in agreement with those obtained for our model LT/SF galaxies.

\subsection{Galaxy size and DM halo properties}
Recently, several studies focused on the relation between galactic and DM halo sizes. 
This is motivated by theoretical work indicating a strong correlation between these quantities: $R_{1/2}\propto R_{200} \lambda$ \citep{mo1998DI}. 
While  $R_{1/2}$ can be measured from observed data, $R_{200}$ and $\lambda$ need to be inferred, in observational studies, using e.g. abundance matching techniques..
%------- SMHM relation evolution for different models; ang_mom/j_integration/redshift/plot_m200_ms_fits.py (fits for X17 made using plot_final_MH_MS.py)
\begin{figure}
    \includegraphics[trim=1cm 0.cm 1.5cm 1cm, clip, width = 0.99\linewidth]{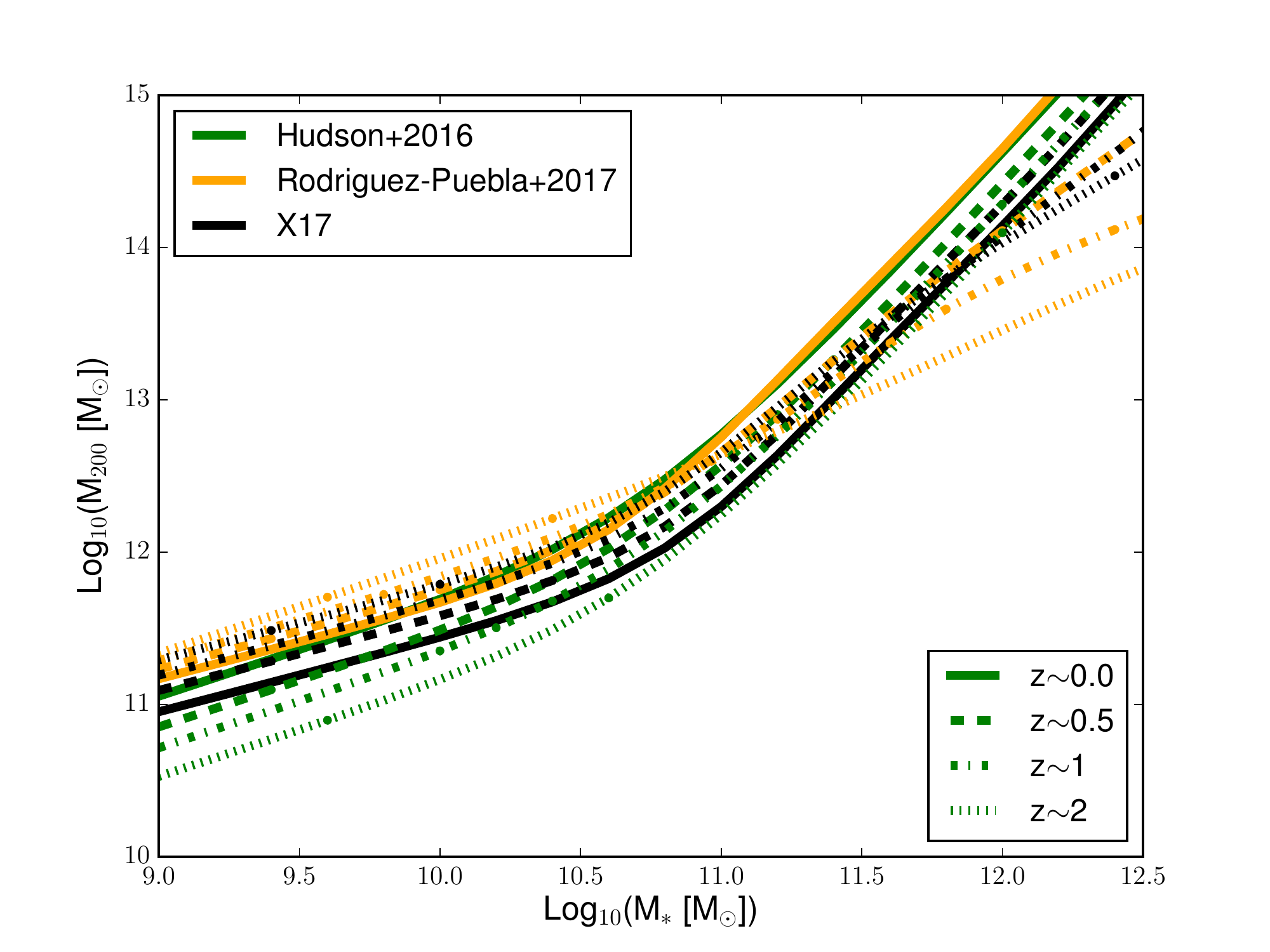}
    \caption{The halo--stellar mass relations (different colors) considered in this work, and their evolution with redshift (different line styles). 
    The relation by\citet{hudson2015SMHM} is shown in green and that by  \citet{rodriguez-puebla2017SMHM} in orange. 
    Results obtained fitting the \citet[][X17]{xie2017sam} model are shown in black.
    }
    \label{fig:mh_ms_comparison}
\end{figure}

Our model galaxies have a $R_{1/2}$--$R_{200}$ relation compatible with those estimated  by \citet{kravtsov2013r50_r200},  and by \citet{huang2017r50_r200}. 
Our model Q and ET galaxies tend to be too small at large values of $R_{200}$, which is probably related to our underestimation of the size of massive ET galaxies.
Our model also predicts little evolution as a function of redshift, in agreement with observational estimates.

We also considered the relation between $R_{1/2}/(R_{200}\lambda)$ and $M_*$.
In this case, we qualitatively compare our results to those by \citet{somerville2018r50_r200}, who have used different estimators for the halo virial mass and radius. 
We find that $R_{1/2}/(R_{200}\lambda)$ is about constant for SF and LT galaxies with stellar masses $M_*<10^{10.5}\,{\rm M_{\sun}}$, and decreases at larger masses. 
The ratio is somewhat higher at increasing redshift, and becomes $\sim1$ at $z\sim2$. 
Q galaxies have a $R_{1/2}/(R_{200}\lambda)$ ratio similar to that of SF galaxies at low/intermediate masses at low redshift, while it decreases at higher redshifts and larger masses.
ET galaxies have a much smaller size at fixed $R_{200}\lambda$.

The behavior found for massive  galaxies can be ascribed to AGN feedback, that freezes the $R_{1/2}$ at early times. 
In this way, the cold gas disk retains the $j_h$ acquired at the last significant cooling event, while the DM halo continues growing its specific angular momentum with cosmic time.

%------- R50-R200 relation at z=0 for SMHM; ang_mom/j_integration/redshift/plot_SMHM_multiple.py
\begin{figure}
    \includegraphics[trim=0cm 0.cm 1.5cm 1cm, clip, width = 0.99\linewidth]{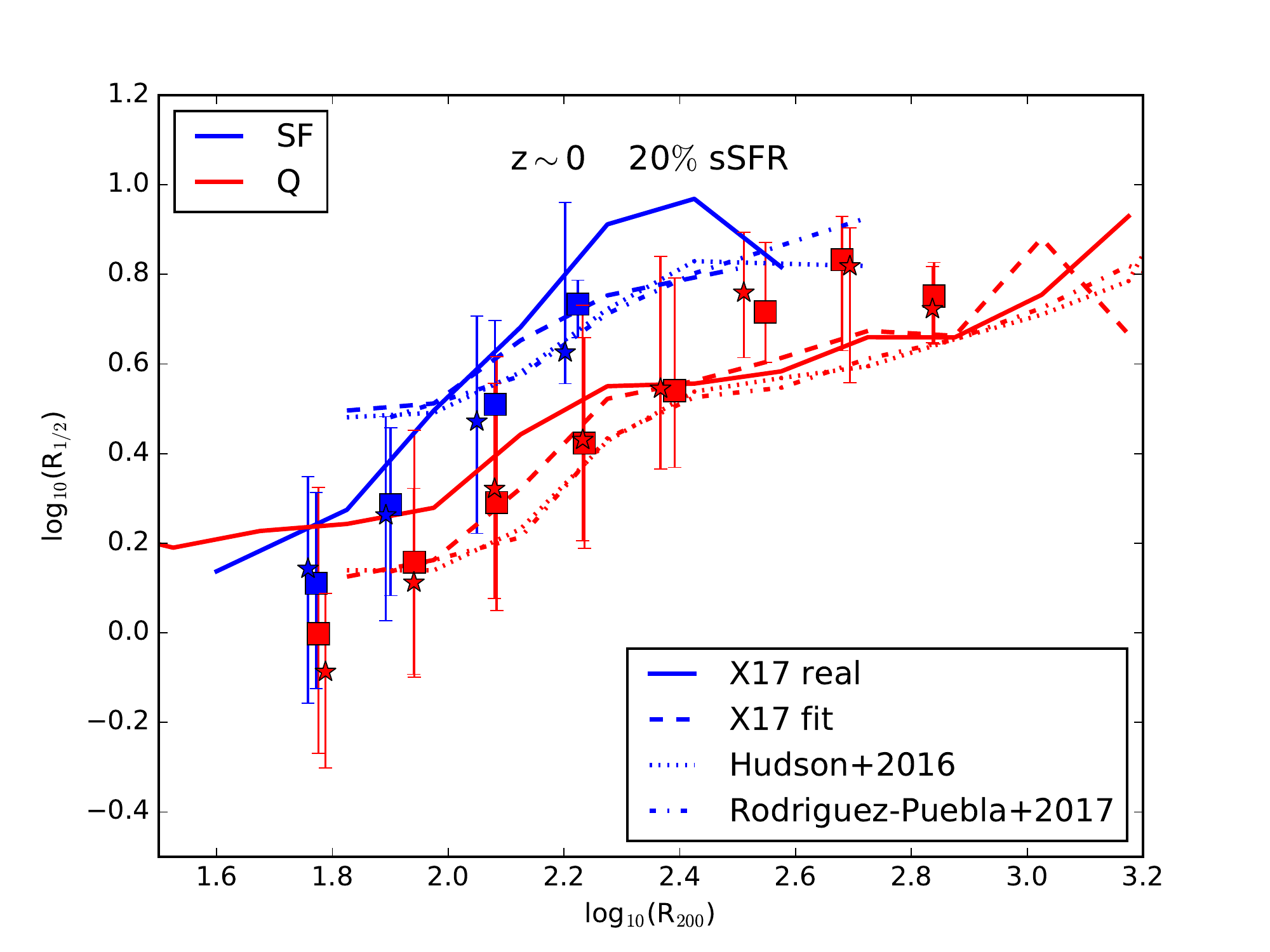}
    \includegraphics[trim=0cm 0.cm 1.5cm 1cm, clip, width = 0.99\linewidth]{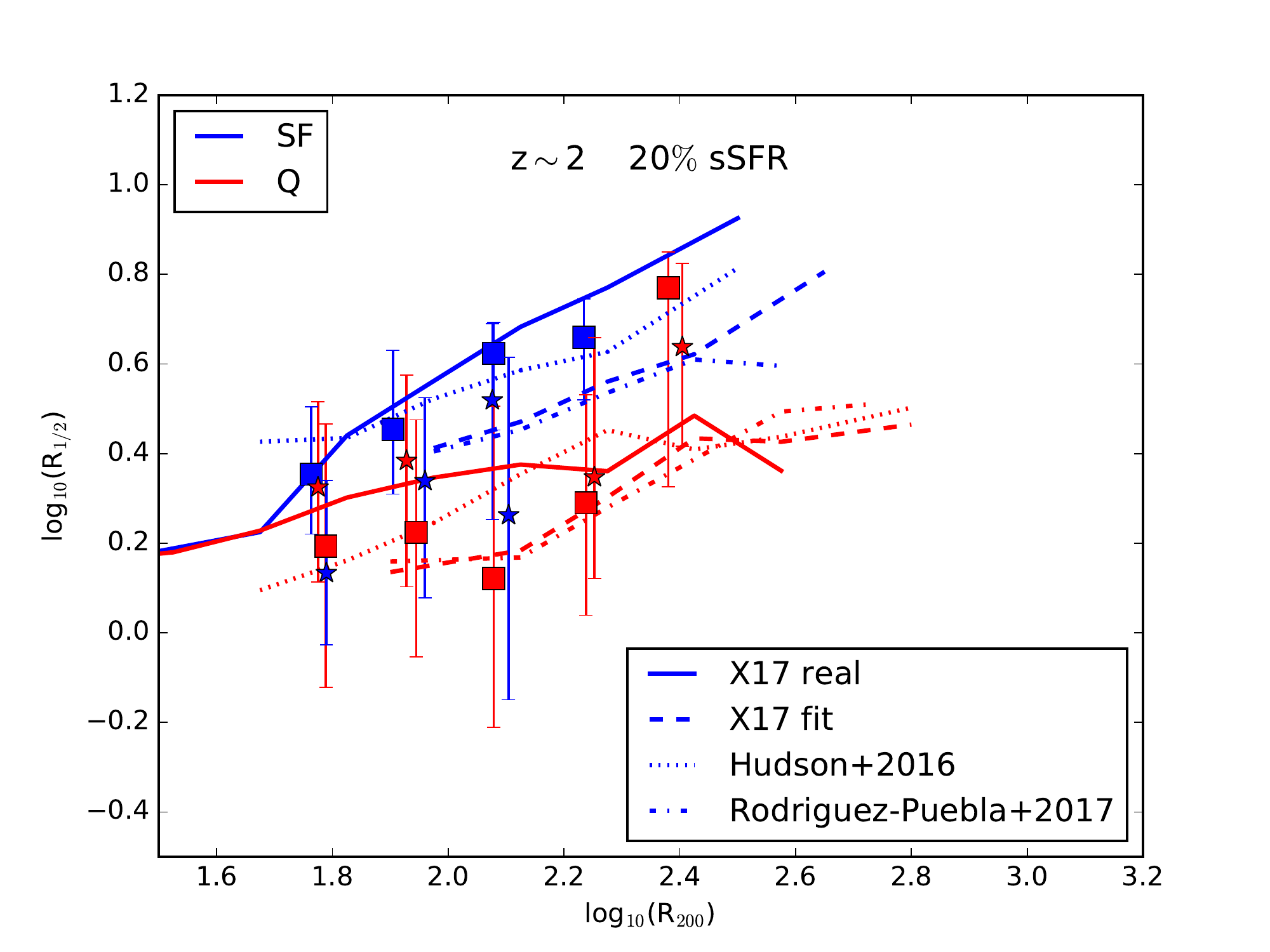}
    \caption{The $R_{1/2}$--$R_{200}$ relations at redshift $z=0$ (top panel) and $z\sim2$ (bottom panel),  obtained using different SMHM relations (different line styles, as indicated in the legend). 
    Results from our model are shown as solid lines.
    Galaxies are divided in LT (blue) and ET (red) selecting the $20\%$ tail distribution in sSFR.
    Observational data are from \citet{huang2017r50_r200}.
    }
    \label{fig:r200_r50_SMHM_z}
\end{figure}

We have analyzed the effect of the abundance matching approximation by ignoring the halo properties of our model galaxies, and assuming  different SMHMs to infer  halo masses and radii.
We assume that SF, Q, LT, ET, central and satellite galaxies have all the same SMHM, and we use three different SMHMs: the one by \citet{hudson2015SMHM}, that by \citet{rodriguez-puebla2017SMHM} and that obtained fitting the SMHM  by model outputs. 
These three SMHM relations are only slightly different, in particular in their normalization (see figure~\ref{fig:mh_ms_comparison}).
The $R_{1/2}$--$R_{200}$ relations inferred are shown in figure~\ref{fig:r200_r50_SMHM_z}, at redshift $z=0$ and at $z\sim2$.
Different styles correspond to different SMHMs, as indicated in the legend. 
We show results only for SF and Q galaxies selected as the 20$\%$ of galaxies with the highest and lowest sSFR, respectively. 

We find that the relations obtained using the abundance matching approach tend to underestimate direct model outputs.
In particular, the normalization of the SMHM relation influences significantly the obtained $R_{200}$. 
In all cases, however, the inferred relations are within the scatter of the observational estimates.

%------------------------------------------------
%------------------CONCLUSIONS-------------------
%------------------------------------------------
\section{Conclusions}
\label{sec:conclusions}
In this work, we have analyzed in detail the evolution of the dynamical properties of galaxies, and their dependence on the hosting DM halo properties.
We have taken advantage of a state-of-the-art semi-analytic model, that includes a sophisticated treatment for metals and energy recycling, an explicit partition of the cold gas into its molecular and atomic components, and a star formation law based on the molecular gas surface density \citep{delucia2014,hirschmann2015,xie2017sam}. 
This model includes also a treatment for specific angular momentum transfer among galactic components, and for computing sizes of stellar and gaseous disks from their specific angular momenta. 
Bulge sizes are evaluated through energy conservation arguments during mergers and disk instabilities, integrated with a treatment for dissipation during mergers. 
In a previous work, we used this model to study the size--mass and specific angular momentum--mass relations at redshift 0, finding they are well reproduced for LT and ET galaxies selected according to their $B/T$ \citep{zoldan2018size_j}. 
In this study, we extend this work to higher redshift and include other dependencies. 
We modify our previous treatment assuming that the rotational velocity of DM halos does not change after they become substructures of bigger halos. 
This is relevant to correctly follow the evolution of satellite galaxy properties, and leads to a better agreement with observed galaxy sizes of passive galaxies.

The main results of our work are summarized below:
\begin{itemize}
 \item Our predicted size--mass relations for SF and LT galaxies are in good agreement with   observations up to high redshift ($z\sim2$), independently of the selection adopted. 
 \item At low to intermediate stellar masses ($M_*<10^{11}\,{\rm M_{\sun}}$), the size--mass relation obtained for Q galaxies is in fair agreement with observations, up to high redshift. 
 Similar results are obtained for ET galaxies whose bulges formed mainly through mergers.
 The relation for ET galaxies in general shifts below observed data at high redshift. 
 This is due to the increasing importance of disk instability in the formation of bulges in central galaxies at high redshifts. 
 \item The size--mass relation of Q and ET galaxies under-estimates the observed relation for large stellar masses ($M_*>10^{11}\,{\rm M_{\sun}}$). 
 Simple variations of the prescriptions regulating various physical processes cannot ease this tension. 
 We find some improvement by switching off AGN feedback or dissipation during mergers. 
 Both these model variants, however, influence strongly other galactic properties.
 We argue that a proper implementation of quasar driven winds could alleviate this problem.
 \item Compact galaxies in our model have number densities compatible with observational estimates. 
 They are not the  main progenitors of today passive galaxies. 
 Rather, they likely remain compact after quenching, and are merged as satellites of larger systems. 
 $z=0$ passive galaxies have quenched relatively recently. 
 \item The evolution of the specific angular momentum--mass relation is in agreement with observational measurements for SF galaxies. 
 Our model also predicts a moderate evolution for the normalization of the Q and ET relations. 
 These results are consistent with the recent findings based on high resolution  hydrodynamical simulations. 
 \item We investigated the relation between galaxy size and halo size, through the evolution of the $R_{1/2}$--$R_{200}$ relation and of the $R_{1/2}/(R_{200}\lambda)$--$M_*$ relation. 
 We find a good agreement with observational estimates, for SF, LT, Q and ET galaxies, validating theoretical models, predicting $R_{1/2}\propto\lambda R_{200}$. 
 \item We checked the influence of the abundance matching  approach on observational estimates of the $R_{1/2}$ vs $R_{200}$ relation. 
 We find consistent results, but slightly different according to the normalization of the stellar to halo mass relation.
 
\end{itemize}

%------------------------------------------------
%-------------AKNOWLEDGEMENTS--------------------
%------------------------------------------------
\section*{Acknowledgements}
We are grateful to Prof. Fall for useful and constructive comments. 
We acknowledge financial support from INAF though the project FORECaST.

%%%%%%%%%%%%%%%%%%%%%%%%%%%%%%%%%%%%%%%%%%%%%%%%%%

%%%%%%%%%%%%%%%%%%%% REFERENCES %%%%%%%%%%%%%%%%%%

% The best way to enter references is to use BibTeX:

\bibliographystyle{mnras}
\bibliography{biblio_1} % if your bibtex file is called example.bib

%%%%%%%%%%%%%%%%%%%%%%%%%%%%%%%%%%%%%%%%%%%%%%%%%%

%%%%%%%%%%%%%%%%% APPENDICES %%%%%%%%%%%%%%%%%%%%%
% 
\appendix

\section{Size evolution in central galaxies}
\label{appendix:r_ms_centrals}
\begin{figure*}
    \includegraphics[trim=2cm 0.8cm 2.5cm 2cm, clip, width = 0.99\linewidth]{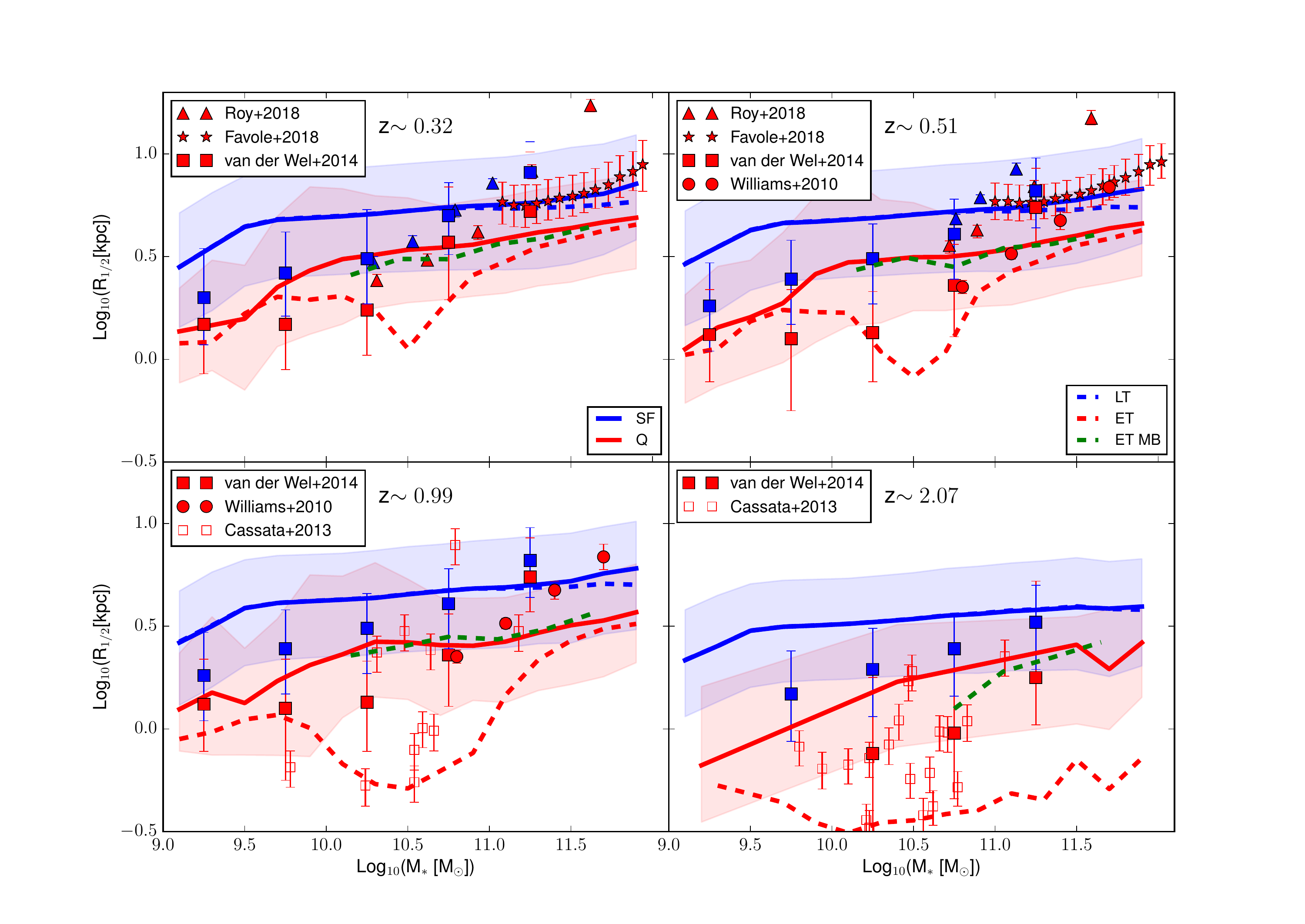}
    \caption{As in figure~\ref{fig:reff_ms_redshift}, but for central galaxies only. }
    \label{fig:reff_ms_redshift_centrals}
\end{figure*}

In figure~\ref{fig:reff_ms_redshift_centrals}, we show the same data as in figure~\ref{fig:reff_ms_redshift}, but for model central galaxies only. 
%------- Reff vs M*  for central galaxies only, at various redshifts; ang_mom/j_integration/c_func/halo_prop/plot_final_reff_ms.py (profiles made with calc_more_jaffe.py)
The size--mass relations obtained for central galaxies at high redshift are very similar to what was found at redshift $z=0$ \citep{zoldan2018size_j}: central galaxies with $B/T>0.5$ have unrealistically small sizes in the stellar mass range $M_*\in[10^{10}-10^{11}]\,{\rm M_{\sun}}$. 
At redshift $z\sim2$, this effect extends to higher masses, as most of these galaxies are formed through disk instabilities.

%%%%%%%%%%%%%%%%%%%%%%%%%%%%%%%%%%%%%%%%%%%%%%%%%%

% Don't change these lines
\bsp	% typesetting comment
\label{lastpage}
\end{document}